\documentclass[extra]{gji}

\usepackage{amsmath,amssymb,mathtools,booktabs,enumitem,relsize,accents}
\usepackage{newtxtext,newtxmath,xcolor}

\newcommand{\avg}[1]{\left\langle#1\right\rangle}
\newcommand{\BPol}{\V B_{\rm Pol}}
\newcommand{\BTor}{\V B_{\rm Tor}}
\newcommand{\curl}{\nabla\times}
\newcommand{\dd}{{\rm d}}
\newcommand{\dr}[1]{\frac{\dd#1}{\dd r}}
\newcommand{\dOmega}{\sin\theta\,\dd\theta\,\dd\phi}
\newcommand{\Eq}[1]{\eqref{#1}}

\newcommand{\FBt}{F_{\!\protect\accentset{\mbox{\normalsize .}}{B}}}
\newcommand{\Fig}[1]{Fig.\,\ref{#1}}
\newcommand{\Figs}[1]{Figs.\,\ref{#1}}
\newcommand{\FF}[1]{Figure\,\ref{#1}}
\newcommand{\FFs}[1]{Figures\,\ref{#1}}
\newcommand{\glm}{g_l^m}
\newcommand{\hlm}{h_l^m}
\newcommand{\dglm}{{\dot g}_l^m}
\newcommand{\dhlm}{{\dot h}_l^m}
\newcommand{\lpeak}{l_\eta}
\newcommand{\ltau}{l_\tau}
\newcommand{\lhi}{l_{\rm hi}}
\newcommand{\llow}{l_{\rm lo}}
\newcommand{\Plm}{P_l^m}
\newcommand{\pp}[2]{\frac{\partial#1}{\partial#2}}
\newcommand{\Pol}{\mathcal P}
\newcommand{\qG}{q_{lm}^{\mathsmaller G}}
\newcommand{\sG}{s_{lm}^{\mathsmaller G}}
\newcommand{\tG}{t_{lm}^{\mathsmaller G}}
\newcommand{\Rsv}{R_{\rm sv}}
\newcommand{\rE}{r_{\text{\fontsize{6}{\baselineskip}\selectfont E}}}
\newcommand{\rin}{r_{\rm i}}
\newcommand{\rout}{r_{\rm o}}
\newcommand{\sect}[1]{section\,\ref{#1}}
\newcommand{\Sect}[1]{Section\,\ref{#1}}
\newcommand{\tausv}{\tau_{\rm sv}}
\newcommand{\V}[1]{\boldsymbol{#1}}
\newcommand{\VPhilm}{\V\Phi_l^m}
\newcommand{\VPsilm}{\V\Psi_l^m}
\newcommand{\VYlm}{\V Y_l^m}
\newcommand{\Ylm}{Y_l^m}
\newcommand{\Tor}{\mathcal T}
\newcommand{\unit}[1]{\V{\hat #1}}
\newcommand{\uh}{\V u_{\rm h}}
\newcommand{\sumlm}{\sum_{l=1}^\infty \sum_{m=-l}^l}
\newcommand{\tauPol}{\tau_{\rm Pol}}
\newcommand{\tauTor}{\tau_{\rm Tor}}

\definecolor{darkred}{rgb}{0.8,0,0}
\definecolor{black}{rgb}{0,0,0}

\title[\large Scaling of the geomagnetic secular variation time scales]{Scaling of the geomagnetic secular variation time scale}

\author[\large Tsang and Jones]{Yue-Kin Tsang$^1$ and Chris A. Jones$^2$ \\
$^1$School of Mathematics, Statistics and Physics, Newcastle University, Newcastle upon Tyne, NE1 7RU, United Kingdom. E-mail: yue-kin.tsang@newcastle.ac.uk \\
$^2$Department of Applied Mathematics, University of Leeds, Leeds, LS2 9JT, United Kingdom}

\date{\today}
\pagerange{\pageref{firstpage}--\pageref{lastpage}}
\volume{000}
\pubyear{2023}

\begin{document}

\label{firstpage}

\maketitle

\begin{summary}
The ratio of the magnetic power spectrum and the secular variation spectrum measured at the Earth's surface provides a time scale $\tausv(l)$ as a function of spherical harmonic degree $l$. $\tausv$ is often assumed to be representative of time scales related to the dynamo inside the outer core and its scaling with $l$ is debated. To assess the validity of this surmise and to study the time variation of the geomagnetic field $\V{\dot B}$ inside the outer core, we introduce a magnetic time-scale spectrum $\tau(l,r)$ that is valid for all radius $r$ above the inner core and reduces to the usual $\tausv$ at and above the core--mantle boundary (CMB). We study $\tau$ in a numerical geodynamo model. At the CMB, we find that $\tau \sim l^{-1}$ is valid at both the large and small scales, in agreement with previous numerical studies on $\tausv$. Just below the CMB, the scaling undergoes a sharp transition at small $l$. Consequently, in the interior of the outer core, $\tau$ exhibits different scaling at the large and small scales, specifically, the scaling of $\tau$ becomes shallower than $l^{-1}$ at small $l$. We find that this transition at the large scales stems from the fact that the horizontal components of the magnetic field evolve faster than the radial component in the interior. In contrast, the magnetic field at the CMB must match onto a potential field, hence the dynamics of the radial and horizontal magnetic fields are tied together. The upshot is $\tausv$ becomes unreliable in estimating time scales inside the outer core. Another question concerning $\tau$ is whether an argument based on the frozen-flux hypothesis can be used to explain its scaling. To investigate this, we analyse the induction equation in the spectral space. We find that away from both boundaries, the magnetic diffusion term is negligible in the power spectrum of $\V{\dot B}$. However, $\V{\dot B}$ is controlled by the radial derivative in the induction term, thus invalidating the frozen-flux argument. Near the CMB, magnetic diffusion starts to affect $\V{\dot B}$ rendering the frozen-flux hypothesis inapplicable. We also examine the effects of different velocity boundary conditions and find that the above results apply for both no-slip and stress-free conditions at the CMB.
\end{summary}


\section{Introduction}
\label{sec:intro}

The time variation of the Earth's magnetic field originates from many different processes and covers a wide range of time scales. In particular, the secular variation, which ranges from years to centuries, is believed to be primarily related to the magnetohydrodynamics in the outer core that generates the geomagnetic field. Thus the secular variation observed at the Earth's surface is often used to infer properties of the dynamo in the planetary interior \citep[e.g.][]{Christensen04}. However, the form of the magnetic field above the core--mantle boundary (CMB) is constrained by the magnetic boundary conditions at the CMB. So to what extent is the time dependence of the magnetic field inside the outer core revealed by the secular variation at the surface? What is the role played by the boundary conditions at the CMB? In this paper, we investigate these issues in terms of the power spectra of the internal magnetic field $\V B$ and its time derivative $\dot{\V B}=\partial\V B/\partial t$.

Although geomagnetic secular variation has been recorded and studied since the 16th century or earlier \cite[]{Jackson15}, high quality data has only been recently available thanks to the satellite missions of {\O}rsted \cite[]{Neubert01}, CHAMP \cite[]{Reigber02}, SAC-C \cite[]{Colomb04} and Swarm \cite[]{Olsen13} that launched between 1999 and 2013. The broad global coverage and long time span of the magnetic measurement provided by these missions, supplemented by observations from ground-based stations, make it possible to construct accurate time-dependent models of the magnetic field at the Earth's surface \citep[e.g.,][]{Lesur10, Finlay20, Fournier21, Hammer21}. These models provide the Gauss coefficients $\glm(t)$ and $\hlm(t)$ as well as their time derivatives $\dglm(t)$ and $\dhlm(t)$ as a function of time $t$. The Gauss coefficients completely specify $\V B$ above the CMB through the relation $\V B=-\nabla V$ where the scalar potential $V$ is given by
\vspace{-0.12cm}
\begin{multline}
V(r,\theta,\phi,t) = \rE \sum_{l=1}^\infty \sum_{m=0}^l \left(\frac{\rE}r \right)^{l+1}\Plm(\cos\theta) \\
\times [\glm(t) \cos m\phi + \hlm(t) \sin m\phi], \quad r \geqslant \rout.
\label{vv}
\end{multline}
Here, $(r,\theta,\phi)$ are the standard spherical coordinates based on the rotation axis, $\rE$ is the mean radius of the Earth and $\rout$ is the radius of the outer core. $\Plm$ are the Schmidt semi-normalised associated Legendre polynomials of degree $l$ and order $m$. It follows from \Eq{vv} that $\dot{\V B}$ above the outer core is fully described by $\dglm(t)$ and $\dhlm(t)$.

Magnetic field structures of different spatial scales may vary on different time scales. This motivates using a spectral approach in the study of secular variation. In terms of the Gauss coefficients, the magnetic power spectrum, or the Lowes--Mauersberger spectrum \cite[]{Mauersberger56,Lowes66}, is defined as
\begin{equation}
R(l,r,t) = \left(\frac{\rE}r \right)^{2l+4}(l+1) \sum_{m=0}^l\left[ (\glm)^2 + (\hlm)^2 \right], \quad r \geqslant \rout.
\label{Rl1}
\end{equation}
By analogy, a secular variation spectrum is defined using the time derivative of the Gauss coefficients \cite[]{Lowes74},
\begin{equation}
\Rsv(l,r,t) = \left(\frac{\rE}r \right)^{2l+4}(l+1) \sum_{m=0}^l\left[ (\dglm)^2 + (\dhlm)^2 \right], \quad r \geqslant \rout.
\label{Rsv}
\end{equation}
From $R$ and $\Rsv$, a spectrum of time scale can be constructed \cite[]{Booker69},
\begin{equation}
\tausv(l,t) = \sqrt{\frac{R}{\Rsv}}, \quad r \geqslant \rout.
\label{tausvdef}
\end{equation}
Note that $\tausv$ is independent of $r$.

The spectrum $\Rsv$ has been obtained for magnetic field models built from satellite data. A review is given by \cite{Gillet10} for earlier generations of these models. More recent results are reported in, for example, \cite{Holme11} and \cite{Finlay20}. Observational errors and influence of the lithospheric field mean that $\Rsv$ can be reliably computed only up to about $l=17$ or so \cite[]{Finlay20}. It is found that in this range of $l$, $\Rsv$ increases with $l$ in a manner that is compatible to $l(l+1)$ \cite[]{Holme11}, or possibly $l^2$. Most likely, $\Rsv$ reaches a maximum at some $l > 17$ before decreasing at large $l$. The full shape of $\Rsv$ is currently not known from observational data and the location of its maximum is an open question.

Geomagnetic secular variation has also been studied using numerical models by, for example, \cite{Christensen03}, \cite{Christensen04} and \cite{Lhuillier11b}. A common result of these studies is that $\tausv$ follows the power law
\begin{equation}
\tausv = \tau_* \cdot l^{-\gamma}
\label{taupwrlaw}
\end{equation}
with the scaling exponent $\gamma=1$. More recent simulations \cite[]{Christensen12,Bouligand16} obtained the same scaling and \cite{Amit18} demonstrated numerically that $\gamma=1$ holds separately for the equatorial symmetric and antisymmetric parts of the magnetic field. On the other hand, analysis using satellite data gives diverse values of $\gamma$. Several works found values in the range $1.32 \leqslant \gamma \leqslant 1.45$ \cite[]{Holme06, Olsen06, Lesur08, Hulot10, Holme11} but $\gamma=1$ was also reported \cite[]{Lhuillier11b, Christensen12, Amit18}. We note that in some of the previous works, $\tausv$ is calculated for a given epoch while in others, its long-time average is considered. This may partly account for the disparities in the observed value of $\gamma$, as discussed in \cite{Lhuillier11b} and \cite{Bouligand16}.

Theoretically, \cite{Lowes74} proposed that $\Rsv \sim l(2l+1) R$. This implies the scaling $\tausv \sim 1/\sqrt{l(2l+1)}$ which is almost indistinguishable from $\tausv \sim l^{-1}$. An argument can also be made for $\gamma=1$ if one invokes the frozen-flux hypothesis and then further assumes the horizontal derivative scales as $\nabla_{\rm h} \sim l$ \cite[]{Holme06,Christensen12}. This frozen-flux argument is meant only for the radial component $B_r$ of the magnetic field and $\tausv$ in \Eq{tausvdef} is indeed fully specified by $B_r$ alone because $\V B$ is constrained to match onto a potential field at the CMB. There is no such constraint in the interior of the outer core. So it is interesting to see if the frozen-flux argument can be used to predict time scales of the dynamo in the interior, given magnetic diffusion is likely to be very small there. Another consequence of the magnetic boundary condition is magnetic diffusion becomes important in a boundary layer below the CMB \cite[]{Jault91,Braginsky93}. Then, is it valid to use the frozen-flux hypothesis to explain the scaling of $\tausv$ above the CMB? Furthermore, if the no-slip condition is used for the velocity, an Ekman--Hartmann layer may develop. Does it play any role here? What is the dominant balance between different effects inside the CMB boundary layer and how does it lead to the scaling of $\tausv$?

In addition to the scaling exponent $\gamma$, \Eq{taupwrlaw} also introduces a time scale $\tau_*$. Being interpreted as a typical secular variation time scale, $\tau_*$ was used to deduce the magnetic Reynolds number inside the outer core \cite[]{Christensen04}, to estimate the predictability of the geodynamo \cite[]{Lhuillier11a} and to rescale the time axis for the interpretation of simulation results \cite[]{Bouligand16}. This leads to the important question of whether $\tau_*$ can generally be considered as a typical time scale for the time variation of the magnetic field inside the outer core.

As discussed above, all previous results are based on the Gauss coefficients and refer to the behaviour of the magnetic field outside the core. In this paper, we introduce a magnetic energy spectrum $F(l,r)$, a magnetic time-variation spectrum $\FBt(l,r)$ and a magnetic time-scale spectrum $\tau(l,r)$ that are defined for all regions above the inner core. We examine these spectra at different depth $r$, including the CMB boundary layer, using a numerical geodynamo model in order to shed some light on the questions raised in this section. A main result is that for the magnetic field components with smaller $l$, the scaling of the time scales with $l$ at the core surface is different from that in the interior of the outer core. So it is generally unreliable to infer typical magnetic time scales inside the outer core from surface observations.

\section{Spectra for magnetic field time variation inside the outer core}

We start by reviewing the magnetic energy spectrum previously used by \cite{Tsang20} in their study of the Jovian dynamo. For a given magnetic field $\V B(r,\theta,\phi,t)$, the magnetic energy spectrum $F(l,r,t)$ at radius $r$ and time $t$ is defined by the relation
\begin{equation}
\sum_{l=1}^\infty F(l,r,t) \equiv \frac{1}{4\pi} \oint |\V B(r,\theta,\phi,t)|^2 \dOmega.
\label{Fdef}
\end{equation}
Unlike the Lowes--Mauersberger spectrum $R(l,r,t)$ in \Eq{Rl1}, $F(l,r,t)$ is defined for all $r$ greater than the radius of the inner core. The expression on the right of \Eq{Fdef} is proportional to the average magnetic energy over a spherical surface of radius $r$. Loosely speaking, $F(l,r,t)$ indicates the amount of magnetic energy residing in the spatial scale characterised by the spherical harmonic degree $l$. We now expand $\V B$ in terms of a basis set of vector spherical harmonics $\big\{ \VYlm(\theta,\phi), \VPsilm(\theta,\phi), \VPhilm(\theta,\phi) \big\}$ (defined in Appendix\,\ref{vsh}):
\begin{equation}
\V B = \sumlm \big[ q_{lm}(r,t) \VYlm + s_{lm}(r,t) \VPsilm + t_{lm}(r,t) \VPhilm \big].
\label{Bvsh}
\end{equation}
From \Eq{Brtpqst}, $q_{lm}$ is the coefficient in the spectral expansion of $B_r$ while the pair $(s_{lm},t_{lm})$ is related to $B_\theta$ and $B_\phi$. Substituting \Eq{Bvsh} into \Eq{Fdef} yields
\begin{equation}
F(l,r,t) = \frac{1}{(2l+1)}\sum_{m=0}^l \big(|q_{lm}|^2 + |s_{lm}|^2 + |t_{lm}|^2\big) (4-3\delta_{m,0}),
\label{Fqst}
\end{equation}
where $\delta_{m,0}$ is the Kronecker delta. For the rest of this paper, we are mostly interested in the time-averaged spectrum $F(l,r) \equiv \avg{F(l,r,t)}_t$, where $\avg{\cdot}_t$ denotes the long-time average over a statistically steady state.

Next, we define the magnetic time-variation spectrum $\FBt(l,r)$. From \Eq{Bvsh}, we have the expansion for the time derivative of the magnetic field:
\begin{equation}
\dot{\V B} = \sumlm \big[ \dot q_{lm}(r,t) \VYlm + \dot s_{lm}(r,t) \VPsilm + \dot t_{lm}(r,t) \VPhilm \big].
\label{Btvsh}
\end{equation}
Then in complete analogue to \Eq{Fdef}, the time-variation spectrum $\FBt$ is defined by
\begin{equation}
\sum_{l=1}^\infty \FBt(l,r,t) \equiv \frac{1}{4\pi} \oint |\dot{\V B}(r,\theta,\phi,t)|^2 \dOmega
\label{FBtdef}
\end{equation}
and its time average is given in terms of the expansion coefficients in \Eq{Btvsh} as
\begin{equation}
\FBt(l,r) = \frac{1}{(2l+1)}\sum_{m=0}^l \avg{ |\dot q_{lm}|^2 + |\dot s_{lm}|^2 + |\dot t_{lm}|^2 }_t (4-3\delta_{m,0}).
\label{FBtqst}
\end{equation}

Finally, equipped with $F$ and $\FBt$, we construct the magnetic time-scale spectrum $\tau(l,r)$ as follows,
\begin{align}
\tau(l,r,t) &\equiv \sqrt{ \frac{F(l,r,t)}{\FBt(l,r,t)} }, \label{tautdef} \\
\tau(l,r) &\equiv \avg{\tau(l,r,t)}_t.
\label{taudef}
\end{align}
We choose to time average the instantaneous $\tau(l,r,t)$ rather than defining $\tau(l,r)$ using the time averages of $F(l,r,t)$ and $\FBt(l,r,t)$. Our results appear to be insensitive to this choice. Note that $\tau$ generally is a function of $r$. We assume the mantle is electrically insulating. Then by definition, $F=R$ and $\FBt=\Rsv$ above the CMB and it follows that $\tau=\tausv$ and becomes independent of $r$ for $r \geqslant \rout$.

\section{The numerical model}

We now describe the numerical geodynamo model used in the present study. It consists of a spherical shell of inner radius $\rin$ and outer radius $\rout$ and the radius ratio is $\rin/\rout=0.35$. The electrically conducting Boussinesq fluid that fills the shell has density $\rho(\V r,t)$, kinematic viscosity $\nu$, magnetic diffusivity $\eta$ and is rotating at an angular speed $\Omega$ about the $z$-axis. We consider the situation when convection in the fluid is driven by compositional buoyancy due to light elements being released from the inner core boundary. Consistent with the Boussinesq approximation, the gravity $\V g(r) = -(g_{\rm o}/\rout)r \unit r$ with $g_{\rm o}$ being the magnitude of $\V g$ at the CMB. The density is given by $\rho = \rho_0 [1 - (\Xi - \Xi_0)]$ where $\Xi(\V r,t)$ is the mass fraction of light elements and $\Xi_0(t)$ its volume average. The constant $\rho_0$ is the density when $\Xi=\Xi_0$. Let $\xi(\V r,t) = \Xi - \Xi_0$, then the equations governing $\V B$, $\xi$ and the velocity $\V u$ are:
\begin{subequations}
\begin{align}
\pp{\V u}t + (\V u \cdot \nabla) \V u &+ 2 \Omega \unit z \times \V u \nonumber \\
= -\frac1{\rho_0} \nabla p' &+ \frac{g_{\rm o}}{\rout} \xi r \unit r + \frac1{\rho_0\mu_0}(\nabla\times {\bf B})\times {\bf B} + \nu \nabla^2 \V u, 
\label{u_eq} \\
\pp{\V B}t  &= \nabla \times (\V u \times \V B) + \eta \nabla^2 \V B, \\
\pp{\xi}t + \V u \cdot \nabla \xi &= \kappa \nabla^2 \xi - \frac{S}{\rho_0},
\label{C_eq} \\
\nabla \cdot \V u &= 0, \\
\nabla \cdot \V B &= 0.
\end{align}
\label{model}%
\end{subequations}
In \Eq{u_eq}, $p'$ is a modified pressure and $\mu_0=4\pi \times 10^{-7}$\,H/m is the permeability of free space. In \Eq{C_eq}, $\kappa$ is the diffusivity of the light elements and the constant $S>0$ is a sink to $\xi$. Physically, $S=\rho_0\partial\Xi_0/\partial t$ is the secular rate of increase of $\Xi_0$ due to the homogenisation of light elements into its surrounding.

At the inner boundary, $\V u$ satisfies the no-slip condition. At the outer boundary, we shall consider both the cases of a no-slip and a stress-free condition for $\V u$. We assume it is electrically insulating outside the spherical shell, hence $\V B$ matches onto a potential field at both boundaries. A no-flux condition is employed for $\xi$ at the outer boundary: $\partial \xi/\partial r|_{\rout}=0$. The boundary condition for $\xi$ at the inner boundary is set by the fact that, by definition, the volume average of $\xi$ vanishes. Then if we assume the release of light elements is uniform over the inner boundary, integrating \Eq{C_eq} over the domain gives
\begin{equation}
\pp{\xi}r\bigg|_{r_i} = -\frac{S}{3\kappa\rho_0} \frac{\rout^3 - \rin^3}{\rin^2}.
\end{equation}
Additionally, $\oint \xi \dOmega=0$ is imposed at $r=\rout$ to fix an arbitrary constant in $\xi$ that would be present if flux conditions are used exclusively.

To non-dimensionlise \Eq{model}, we use the shell thickness $d=\rout-\rin$ as the unit of length and $d^2/\eta$ as the unit of time. The units for the fields of $\V u$, $\V B$, $\xi$ and $p'$ are $\eta/d$, $\sqrt{\Omega\rho_0\eta\mu_0}$, $Sd^2/\rho_0\eta$ and $\Omega\rho_0\eta$ respectively. This gives the following non-dimensional form of \Eq{model}:
\begin{subequations}
\begin{align}
\pp{\V u}t + (\V u \cdot \nabla) \V u + 2\frac{Pm}{Ek} & \unit z \times {\bf u} \nonumber \\
= -\frac{Pm}{Ek} \nabla p' + \frac{Ra Pm^2}{Pr} & \xi r \unit r
  + \frac{Pm}{Ek} (\nabla\times {\bf B})\times {\bf B} + Pm \nabla^2 \V u, \\
\pp{\V B}t &= \nabla \times ({\V u} \times {\V B}) + \nabla^2 \V B, 
\label{indeq} \\
\pp{\xi}t + (\V u \cdot \nabla) \xi &= \frac{Pm}{Pr} \nabla^2 \xi - 1,
\\
\nabla \cdot \V u &= 0, \\
\nabla \cdot \V B &= 0,
\end{align}
\label{ndmodel}%
\end{subequations}
where the non-dimensional numbers are:
\begin{equation}
Ra=\frac{g_{\rm o} S d^6}{\rout \rho_0 \eta \kappa \nu}, \quad Ek = \frac{\nu}{\Omega d^2}, \quad Pr=\frac{\nu}{\kappa}, \quad Pm = \frac{\nu}{\eta}.
\label{nondim_nos}
\end{equation}
In these units, the magnetic Reynolds number $Rm$ is given by the root-mean-squared value of the velocity magnitude $|\V u|$ over the spherical shell.

We solve \Eq{ndmodel} numerically using the pseudo-spectral code developed in \cite{Willis07}. We carried out four simulations with the parameters given in Table\,\ref{para_tbl}.
\begin{table}
{\centering
\begin{tabular}{ccccccc}
\hline\\[-0.4cm]
run & $Ra$ & $Ek$ & $Rm$ & $Pr$ & $Pm$\\
\hline
1 & $6.75 \times 10^8$ & $2.5 \times 10^{-5}$ & 504 & 1 & 2.5 & no-slip \\ 
2 & $6.75 \times 10^8$ & $2.5 \times 10^{-5}$ & 702 & 1 & 2.5 & stress-free \\ 
3 & $1.625 \times 10^8$ & $ 1 \times 10^{-4}$ & 474 & 1 & 2.5 & no-slip \\ 
4 & $1.875 \times 10^8$ & $2.5 \times 10^{-5}$ & 260 & 1 & 2.5 & no-slip \\
\hline
\end{tabular}\par}
\caption{Parameter values used in the four simulations considered in this study. The parameters are defined in \Eq{nondim_nos} and the text below it. The last column shows the boundary condition for the velocity at the outer boundary.}
\label{para_tbl}
\end{table}
Each of these simulations produces a dipolar magnetic field that is categorised as ``Earth-like'' in \cite{Christensen10}. We use 162 to 168 grid points in the radial direction and the maximum degree and order in the spherical harmonics expansion are both between 168 and 176. Typical time-step used in the simulations is $\delta t=10^{-7}$. Time derivatives such as $\dot q_{lm}$ in \Eq{Btvsh} are computed using the forward difference formula
\begin{equation}
\dot q_{lm}(t) = \frac{q_{lm}(t+\Delta t) - q_{lm}(t)}{\Delta t}
\end{equation}
with $\Delta t = 5 \times 10^{-9}$. After the system reaches a statistically steady state, time averages are calculated over about 100 to 150 snapshots spanning about 2 magnetic diffusion times. We first focus on run\,1 which uses the no-slip condition at the outer boundary. Discussion on run\,2 with stress-free condition at $r=\rout$ is postponed until Section~\ref{sec:sf}. Then in \sect{EkRm}, we compare these results with run\,3 which has a larger $Ek$ and run\,4 which has a smaller $Rm$. 

\begin{figure}
\centering
\includegraphics[width=0.45\textwidth]{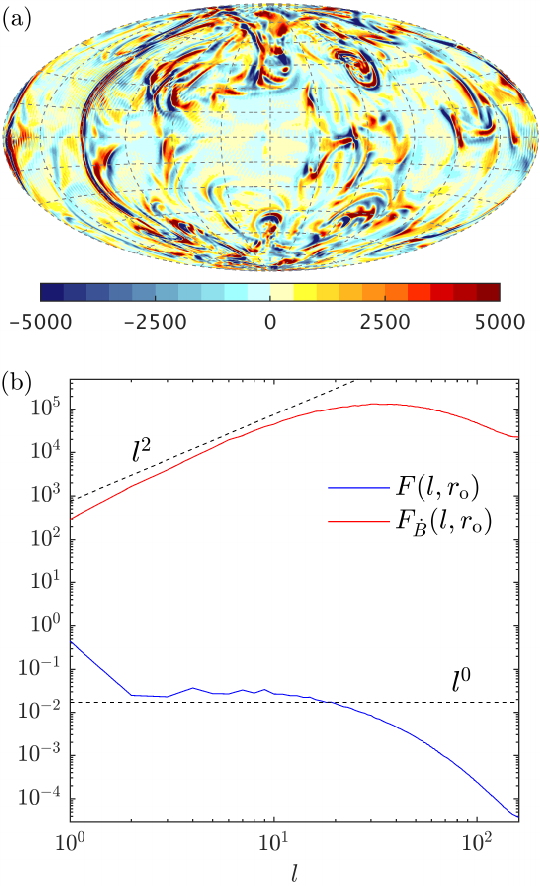}
\caption{For run\,1 in Table\,\ref{para_tbl}: (a) a snapshot of the time derivative of the radial magnetic field $\dot B_r$; (b) the time-averaged magnetic energy spectrum $F$ in \Eq{Fqst} and the magnetic time-variation spectrum $\FBt$ in \Eq{FBtqst} computed at the CMB $r=\rout$.}
\label{fig:dtBr_Fl}
\end{figure}

From run\,1, \Fig{fig:dtBr_Fl}(a) shows a snapshot of the time derivative of the radial magnetic field $\dot B_r$ at $r=\rout$. \FF{fig:dtBr_Fl}(b) plots on the same graph the two spectra $F$ and $\FBt$ at $r=\rout$.
Focusing on the large scales, we see that, excluding $l=0$, $F(l,\rout)$ is flat to a very good approximation on the log scale, compatible with the ``white source hypothesis'' at the CMB. We also find that $\FBt(l,\rout) \sim l^2$ at the large scales, compatible with satellite observations. We shall come back to these scalings in later sections.

\section{The magnetic time-scale spectrum $\lowercase{\tau(l,r)}$}
\label{sec:tau}

\begin{figure}
\centering
\includegraphics[width=0.45\textwidth]{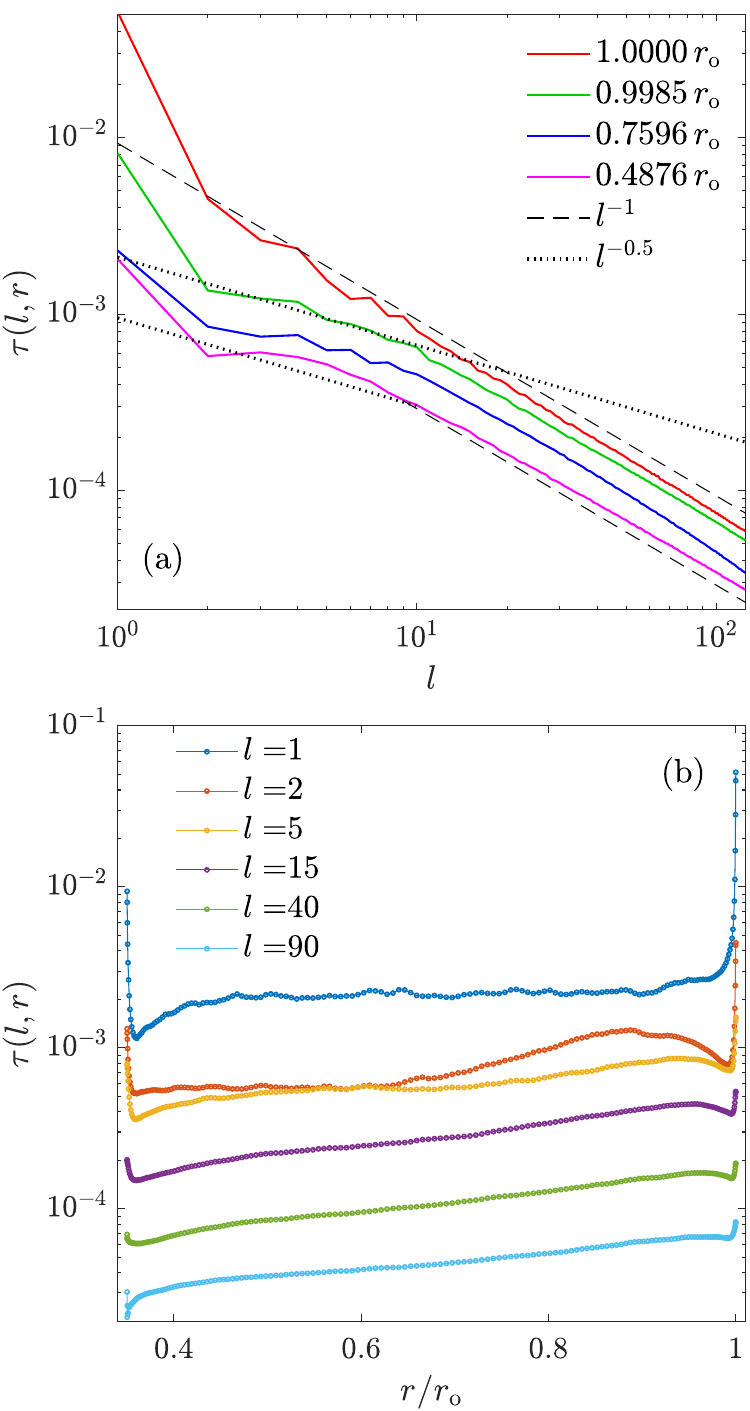}
\caption{(a) Time-averaged magnetic time-scale spectrum $\tau(l,r)$, defined in \Eq{taudef}, at different radius $r$ for run\,1. Here, $\rout$ is the radius of the CMB. This shows the different types of scaling found at the CMB and in the interior. (b) $\tau(l,r)$ as a function of $r$ for selected modes $l$, showing the sharp transition near the CMB.}
\label{fig:tau}
\end{figure}

From our simulation data, we compute the magnetic time-scale spectrum $\tau(l,r)$ as a function of $l$ at different $r$. The results are plotted in \Fig{fig:tau}(a). Generally $\tau(l,r)$ decreases with increasing $l$ for all $r$. If we consider $\tau(l,r)$ as the typical time scale for the temporal variation of magnetic field structures with a spatial scale characterised by $l$, then \Fig{fig:tau}(a) suggests that magnetic field structures of larger spatial scales vary on longer time scales than those of smaller spatial scales. 

Let us now study the scaling of $\tau(l,r)$ with $l$ at different depth inside the outer core. 
At the CMB $r = \rout$, excluding the dipole, we find the clean power-law scaling of
\begin{equation}
\tau \sim l^{-1}, \quad l \geqslant 2.
\label{l1}
\end{equation}
This agrees with results from previous numerical studies but differs from some observational results, as discussed in the introduction. A more intriguing finding of the present work is that $\tau$ behaves quite differently from \Eq{l1} inside the outer core. As $r$ decreases, $\tau$ starts to display different $l$-dependence at the large scales and the small scales. \FF{fig:tau}(a) shows that while $\tau \sim l^{-1}$ still holds for large $l$, $\tau$ becomes shallower than $l^{-1}$ at small $l$, with slope closer to the dotted $l^{-0.5}$ lines than the dashed $l^{-1}$ lines. We denote the value of $l$ at which the slope changes by $\ltau$. We find that, very roughly, $\ltau \approx 13$ in all four simulations listed in Table\,\ref{para_tbl}. $\ltau$ may have a different value for more extreme simulation parameters or in the  Earth. For the main results of the present study, the value of $\ltau$ is not so important, it only marks the boundary of two different scaling regimes.

It is interesting that the change in the scaling of $\tau$ for $l < \ltau$ occurs within a very thin boundary layer under the CMB. We can see in \Fig{fig:tau}(a) that the slope of $\tau$ for $l < \ltau$ at $r=0.9985\rout$ has already changed to $l^{-0.5}$, despite it being $l^{-1}$ at $r=\rout$. \FF{fig:tau}(b) plots $\tau(l,r)$ as a function of $r$ for selected $l$ and illustrates clearly the very rapid changes in the small-$l$ modes, especially the dipole, just beneath the CMB. This sharp slowing down of the large-scale magnetic field within the CMB boundary layer as $r$ increases leads to the simple power-law scaling \Eq{l1} at the surface. For the large-$l$ modes, \Fig{fig:tau}(b) shows that $\tau$ varies only weakly with $r$. Abrupt changes are also found near the inner boundary, however, we shall focus on the interior of the outer core and the CMB, i.e. $r \gg \rin$, for the rest of this paper.

To quantify the results described above, we perform a least-square fit of $\tau(l,r)$ to the power law \Eq{taupwrlaw} at different $r$ and we have done this separately for small and large $l$. \FF{fig:gamma} plots $\gamma(r)$ and $\tau_*(r)$ versus $r$ and also gives the range of $l$ where the fit is performed.
\begin{figure}
\centering
\includegraphics[width=0.45\textwidth]{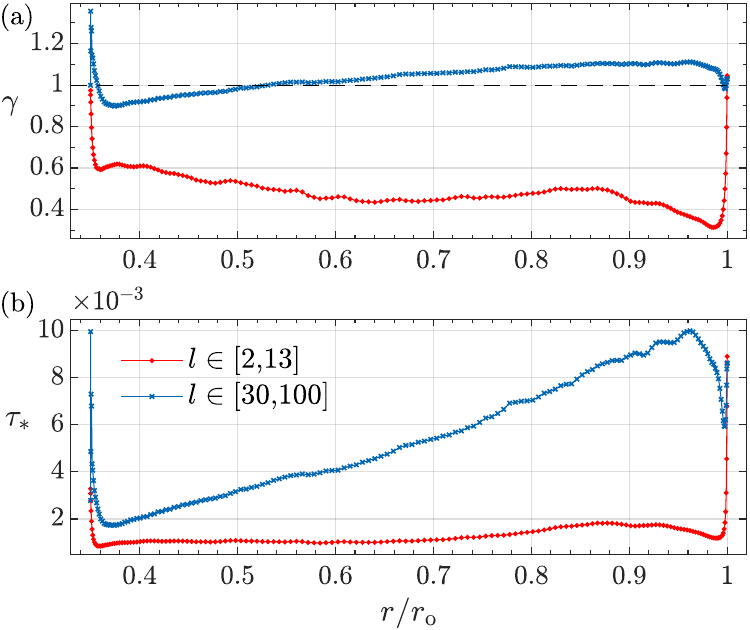}
\caption{The scaling exponent $\gamma$ and the prefactor $\tau_*$ of the power-law fit \Eq{taupwrlaw} to the magnetic time-scale spectrum $\tau(l,r)$ as a function of radius $r$ for run\,1. The least-square fit is performed separately for the large scales (red) and the small scales (blue). The range of spherical harmonic degrees $l$ used for the fit in each case is indicated in the figure.}
\label{fig:gamma}
\end{figure}
For the small scales $l > \ltau$, $\gamma \approx 1$ both at the surface and in the interior, where ``interior'' refers to the region away from both boundaries. For the large scales $l < \ltau$, $\gamma \approx 0.5$ in the interior and then shoots up to $\gamma \approx 1$ at the surface. Thus to a reasonably good approximation, we have the following hybrid scaling in the interior of outer core,
\begin{subequations}
\begin{alignat}{2}
\tau &\sim l^{-0.5}, \quad & 2  \leqslant l & \leqslant 13, \\
\tau &\sim l^{-1},   \quad & 30 \leqslant l & \leqslant 100,
\end{alignat}
\label{gamma}%
\end{subequations}
consistent with the scalings plotted in \Fig{fig:tau}(a).

As discussed in \Sect{sec:intro}, the prefactor $\tau_*(\rout)$ in the power-law fit for the spectrum $\tau(l,\rout)$ at the surface is sometimes taken as a representative time scale of the geomagnetic secular variation. Previous observational studies found $500\,\text{yr} \lesssim \tau_*(\rout) \lesssim 1000\,\text{yr}$. For our simulation, adopting the values of magnetic diffusivity $\eta=0.723$ m$^2$s$^{-1}$ \cite[]{Pozzo12} and shell width \mbox{$d=2.265 \times 10^{6}$ m}, \Fig{fig:gamma}(b) shows that $\tau_*(\rout) \approx 9 \times 10^{-3}$ or 2025\,yr in dimensional unit. We see from \Fig{fig:tau} that this is not representative of $\tau(l,r)$ at any $l$ in the interior where $\tau$ ranges from $10^{-5}$ (2.25\,yr) at the small scales to $10^{-3}$ (225\,yr) at the large scales. For comparison, the dipole at the CMB has a much slower time scale of $\tau(1,\rout) \approx 5 \times 10^{-2}$ or 11250\,yr.

\section{Poloidal and toroidal time scales}
\label{sec:tauPT}

At the CMB, the poloidal part $\BPol$ of the magnetic field matches onto a potential field and the toroidal part $\BTor$ vanishes. It turns out the sharp change in the character of $\tau$ near the CMB described in \Sect{sec:tau} is linked to these changes in $\V B$ dictated by the magnetic boundary condition. To see this, we introduce time-scale spectra associated with $\BPol$ and $\BTor$ and compare them with $\tau$ (which is defined using the full magnetic field $\V B$).

To proceed, let us first define the time-averaged spectra
\begin{align}
F_q(l,r) = \frac{1}{(2l+1)}\sum_{m=0}^l \avg{ |q_{lm}|^2 }_t (4-3\delta_{m,0}), \label{Fqdef} \\
F_{\dot q}(l,r) = \frac{1}{(2l+1)}\sum_{m=0}^l \avg{ |\dot q_{lm}|^2 }_t (4-3\delta_{m,0}),
\label{Fqtdef}
\end{align}
using the decomposition of ${\V B}$ given by \Eq{Bvsh}. We also define $F_s$, $F_{\dot s}$, $F_t$ and $F_{\dot t}$ in a completely analogous manner. Then we see from \Eq{Fqst} and \Eq{FBtqst} that both $F$ and $\FBt$ can be written as a sum of three parts:
\begin{align}
F &= F_q + F_s + F_t, \label{FBFqst} \\
\FBt &= F_{\dot q} + F_{\dot s} + F_{\dot t}.
\label{FBtFqst}
\end{align}
Now via the scalar poloidal potential $\Pol$ and toroidal potential $\Tor$, $\V B$ can be decomposed as:
\begin{subequations}
\begin{align}
\V B &= \BPol + \BTor, \label{BPT1} \\
\BPol &= \curl \curl (\Pol \V r), \label{BPT2} \\
\BTor &= \curl (\Tor \V r), \label{BPT3}
\end{align}
\label{BPT}%
\end{subequations}
where $\V r = r \unit r$. By expanding $\Pol$ and $\Tor$ in terms of the scalar spherical harmonics $\Ylm$ as in \Eq{PTlm}, it can be shown that:
\begin{subequations}
\begin{align}
\BPol &= \sumlm \left( q_{lm} \VYlm + s_{lm} \VPsilm \right), \label{BPqslm} \\
\BTor &= \sumlm t_{lm} \VPhilm,
\label{BTtlm}%
\end{align}
\label{BPTqst}%
\end{subequations}
and $q_{lm}$, $s_{lm}$, $t_{lm}$ are related to the expansion coefficients of $\Pol$ and $\Tor$ by \Eq{qstPlm}. Thus we have decomposed the expansion \Eq{Bvsh} for $\V B$ into a poloidal part and a toroidal part. It is clear from \Eq{BPTqst} or \Eq{qstPlm} that $t_{lm}$ fully describes the toroidal magnetic field while the pair $(q_{lm},s_{lm})$ is related to the poloidal magnetic field.

Following the same logic from \Eq{Bvsh} through \Eq{taudef}, we define a time-scale spectrum associated with $\BPol$ as
\begin{equation}
\tauPol(l,r) = \avg{\! \sqrt{ \frac{F_q + F_s}{F_{\dot q} + F_{\dot s}} }\, }_{\!t}.
\label{tauPoldef}
\end{equation}
Similarly, a time-scale spectrum associated with $\BTor$ is defined as
\begin{equation}
\tauTor(l,r) = \avg{\! \sqrt{ \frac{F_t}{F_{\dot t}} }\, }_{\!t}.
\label{tautordef}
\end{equation}
Finally, we introduce one more time-scale spectrum $\tau_r$ which is relevant to the radial magnetic field $B_r$. Recall from \Eq{Brqlm} that $q_{lm}$ is the spectral coefficient of $B_r$. Using \Eq{VYlmdef} we can write
\begin{equation}
B_r \unit r = \sumlm q_{lm} \VYlm,
\end{equation}
which is of similar form to \Eq{BTtlm}. Therefore we define
\begin{equation}
\tau_r(l,r) = \avg{\! \sqrt{ \frac{F_q}{F_{\dot q}} } \,}_{\!t}.
\label{taurdef}
\end{equation}
One may also want to define, in a similar fashion as above, a time-scale spectrum for $\Pol$. However, as shown in Appendix\,\ref{FP_FT}, such spectrum will be identical to $\tau_r$. Similarly, $\tauTor$ also serves as the time-scale spectrum for $\Tor$.

\FF{fig:tauPT} plots $\tau$, $\tauPol$, $\tauTor$ and $\tau_r$ versus $l$ at three different $r$ representing locations in the interior of the outer core, inside the CMB boundary layer and at the CMB.
\begin{figure*}
\centering
\includegraphics[width=0.97\textwidth]{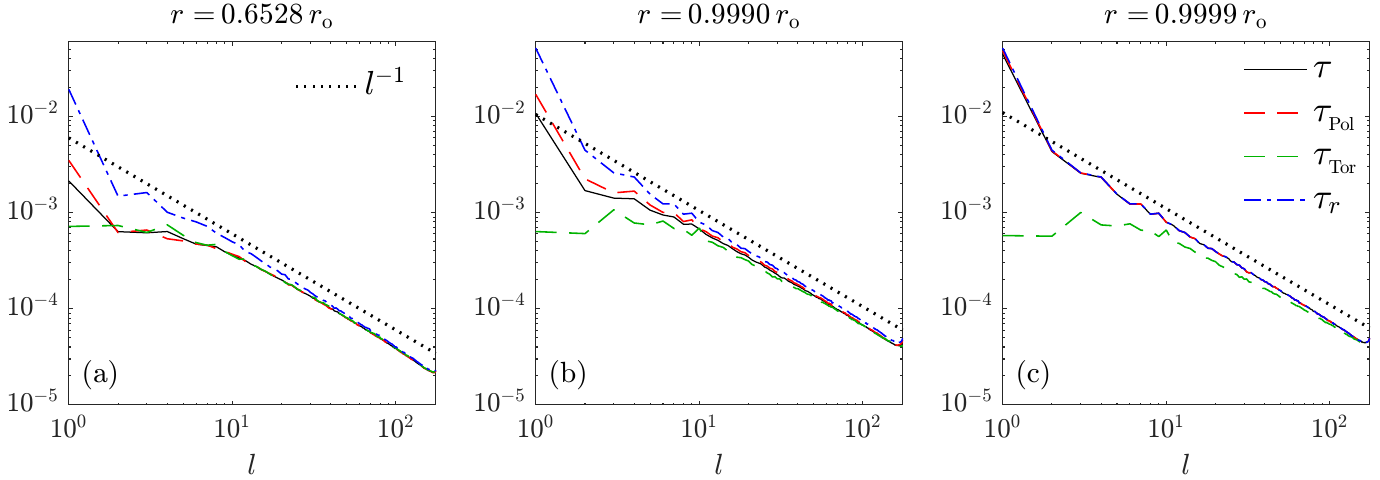}
\caption{Comparison of four (time-averaged) time-variation spectra for run\,1: $\tau(l,r)$ in \Eq{taudef} is associated with the full magnetic field $\V B$, $\tauPol(l,r)$ in \Eq{tauPoldef} is associated with the poloidal magnetic field $\BPol$, $\tauTor(l,r)$ in \Eq{tautordef} is associated with the toroidal magnetic field $\BTor$ and $\tau_r(l,r)$ in \Eq{taurdef} is associated with the radial magnetic field $B_r$. Three different locations are shown: (a) in the interior, (b) inside the CMB boundary layer where transition occurs and (c) just below the CMB. We see that $\tau$ and $\tauPol$ follow each other and change shape as $r \to \rout$ whereas $\tauTor$ and $\tau_r$ maintain their shape for all $r$.}
\label{fig:tauPT}
\end{figure*}
In the interior, \Fig{fig:tauPT}(a) shows that $\tauPol$ and $\tauTor$ essentially overlap (except at $l=1$) and both exhibit the hybrid scaling of \Eq{gamma}. This suggests $\BPol$ and $\BTor$, and consequently $\V B$, all vary with time in a similar fashion and hence
\begin{equation}
\tau \approx \tauPol \approx \tauTor \quad \text{(in the interior).}
\end{equation}

Close to the CMB, \Figs{fig:tauPT}(b) and \ref{fig:tauPT}(c) show that $\tauPol$ becomes steeper at the large scales $l < \ltau$ and in the end follows the single scaling law $\tauPol \sim l^{-1}$ for all $l>1$ at the CMB. On the other hand, $\tauTor$ maintains the hybrid scaling at all $r$. However the shape of $\tauTor$ is irrelevant near the CMB because $\BTor$ diminishes and $\V B$ is dominated by $\BPol$ as $r \to \rout$, so the change of shape in $\tau$ follows closely that of $\tauPol$. And with $\BTor = \V 0$ at the CMB, we have 
\begin{equation}
\quad\qquad \tau = \tauPol \sim l^{-1} \quad \text{(at the CMB).}
\label{tauCMB}
\end{equation}
This is illustrated in \Fig{fig:tauPT}(c) which shows the situation at a single grid point below the CMB
with $r=0.9999\,\rout$. We remark that $\tauTor$ does not tend to zero as $r \to \rout$, though $F_t$ and $F_{\dot t}$ both go to zero as $r \to \rout$. This is why we cannot plot $\tauTor$ at $r=\rout$.

We now discuss the scaling of $\tau_r$. First we note that $\V B$ matching onto a potential field at the CMB implies
\begin{equation}
\pp{\tilde\Pol_{lm}}{r} + \frac{l+1}r \tilde\Pol_{lm} = 0 \quad \text{at $r = \rout$},
\label{PlmBC}
\end{equation}
where $\tilde\Pol_{lm}$ is the coefficients in the spectral expansion \Eq{PTlm} of $\Pol$. Using the condition \Eq{PlmBC} in \Eq{slmPlm} links $s_{lm}$ to $q_{lm}$ at the CMB:
\begin{equation}
s_{lm} = -\sqrt{\frac l{l+1}}\, q_{lm} \quad \text{at $r = \rout$}.
\label{slm_qlm}
\end{equation}
It then follows from \Eq{tauPoldef} and \Eq{taurdef} that $\tau_r=\tauPol$ at $r = \rout$. Hence together with \Eq{tauCMB}, we have $\tau = \tauPol = \tau_r \sim l^{-1}$ at the CMB, as is evident in \Fig{fig:tauPT}(c). Remarkably, \Fig{fig:tauPT} also shows that unlike $\tau$, the scaling $\tau_r \sim l^{-1}$ manifest at $r = \rout$ is actually valid for all $r$:
\begin{equation}
\qquad \tau_r \sim l^{-1} \quad (l \neq 1) \quad \text{for all $r$}.
\label{taur_l1}
\end{equation}
This might lead to the belief that the scaling \Eq{taur_l1} at the surface reflects the properties of $B_r$ in the interior, and furthermore \Eq{taur_l1} might be explained by the frozen-flux argument outlined in the introduction. However, we shall see in \Sect{sec:spec_r} that depending on the velocity boundary condition, the scaling \Eq{taur_l1} found at the surface and in the interior could have different origins.

The above results, summarised in \Fig{fig:tauPT}, suggest that the magnetic boundary condition causes $\tau$ to change shape near the CMB. It eliminates the toroidal contribution to $\tau$ and constrains the poloidal contribution through \Eq{slm_qlm}. As a result, $\tau$ at $r = \rout$ is generally unreliable for inferring time scales of the dynamics in the interior.

By comparing $\tau$ with $\tau_r$ at $l < \ltau$, we can gain some physical understanding of the different scaling of $\tau$ with $l$ at different $r$. In the interior, except for $\nabla \cdot \V B = 0$, $B_r$ and the horizontal components $(B_\theta,B_\phi)$ contribute separately to the dynamics. Since the toroidal potential $\Tor$ is related to $(B_\theta,B_\phi)$, the result of $\tau \approx \tauTor < \tau_r$ in \Fig{fig:tauPT}(a) suggests that in the interior, $(B_\theta,B_\phi)$ evolve faster than $B_r$ and $\dot{\V B}$ is dominated by $(\dot B_\theta,\dot B_\phi)$. At the CMB, the magnetic boundary condition together with $\nabla \cdot \V B = 0$ means $B_r$, $B_\theta$ and $B_\phi$ are related to each other, as reflected in \Eq{slm_qlm} (and $t_{lm}=0$). Consequently, $\tau = \tau_r$ and \Fig{fig:tauPT}(c) shows that $\tau$ then follows the scaling of $\tau_r$. Further details emerge in \sect{sec:specbal} when we investigate the induction equation in the spectral space.

\section{The spectra $F\lowercase{(l,r)}$ and $F_{\dot B}\lowercase{(l,r)}$}

We want to trace the origin of the scaling \Eq{l1} at the CMB and \Eq{gamma} in the interior. In order to do this, we need to first take a closer look at the two spectra $F$ and $\FBt$ that made up $\tau$. Moreover, as discussed in the introduction, the shape of $\FBt$ is in itself an interesting question. Our numerical result for $\FBt$ covers a broad range of $l$ and thus complements observational studies of $\Rsv$ which are limited to fairly small $l$.

\subsection{The magnetic energy spectrum $F(l,r)$}

\begin{figure*}
\centering
\includegraphics[width=\textwidth]{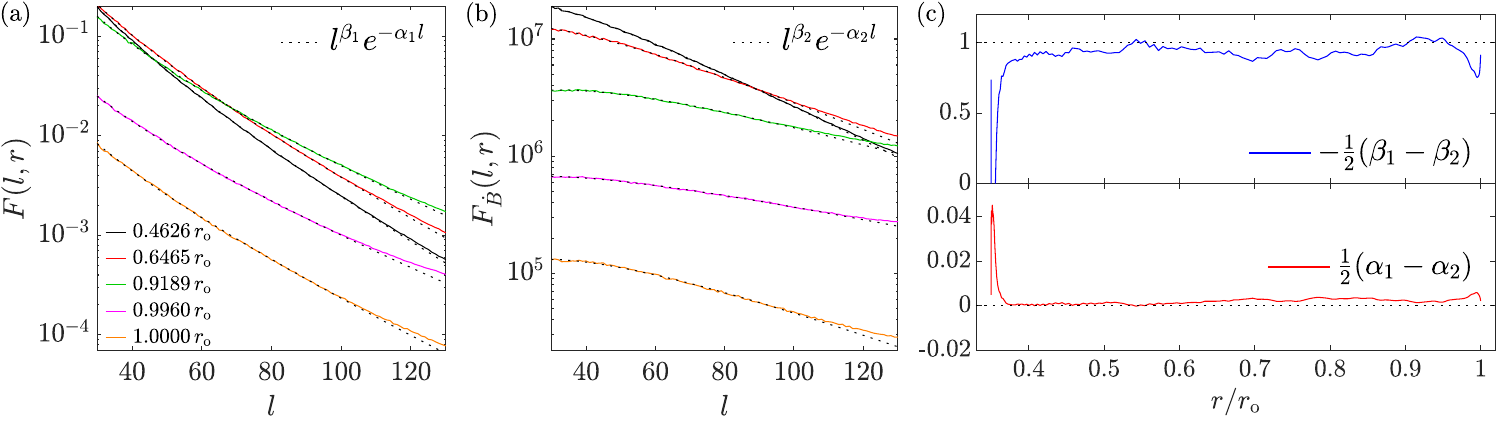}
\caption{Small-scale properties for run\,1: (a) decay of the magnetic energy spectrum $F(l,r)$ in \Eq{Fqst} for different radius $r$. It is well fitted by the form $F \sim l^{\beta_1}\exp(-\alpha_1 l)$; (b) Similar to (a) but for the magnetic time-variation spectrum $\FBt(l,r)$ in \Eq{FBtqst}. The form $\FBt \sim l^{\beta_2}\exp(-\alpha_2 l)$ again fits the data extremely well; (c) The differences between the fitting parameters of $F$ and $\FBt$ showing $\alpha_1 \approx \alpha_2$ and $(\beta_1-\beta_2)/2 \approx -1$ that lead to $\tau \sim l^{-1}$ at large $l$.}
\label{fig:F_hi}
\end{figure*}

\FF{fig:Fl_r} plots the magnetic energy spectrum $F(l,r)$ at different $r$.
\begin{figure}
\centering
\includegraphics[width=0.45\textwidth]{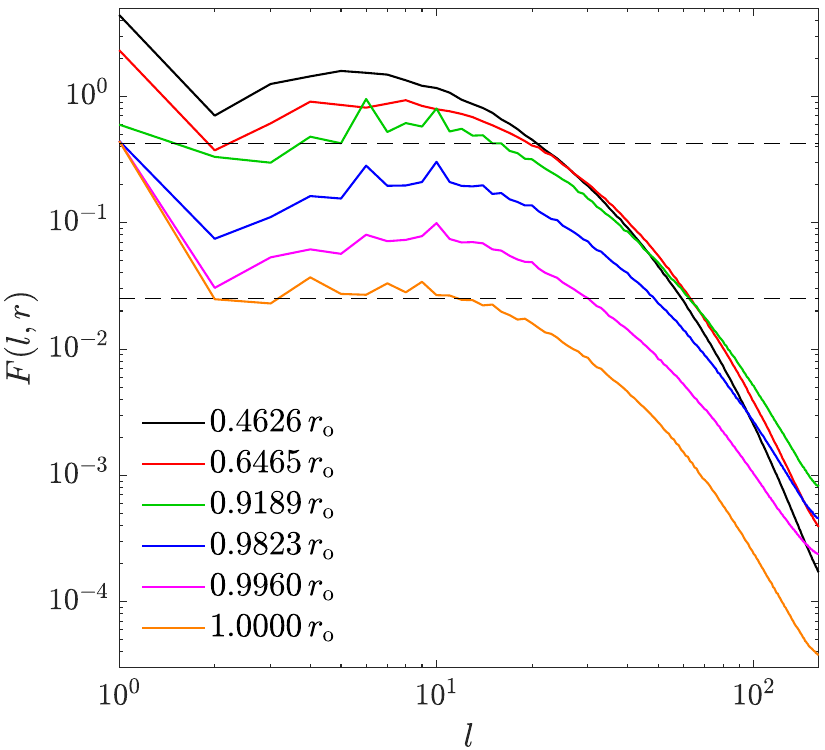}
\caption{Time-averaged magnetic energy spectrum $F(l,r)$, given by \Eq{Fqst}, at different radius $r$ for run\,1. $F(l,r)$ is fairly flat (in logarithmic scale) at the large scales.}
\label{fig:Fl_r}
\end{figure}
We already know from \Fig{fig:dtBr_Fl} that $F(l,r)$ is flat for the small-$l$ modes (excluding $l=1$) at $r=\rout$. Here we see that the \mbox{small-$l$} modes remain fairly flat in the interior. For large $l$, as shown in \Fig{fig:F_hi}(a), $F$ can be very well approximated by an exponential modified by a power law for the whole domain. We thus conclude that for $ \rin \ll r \leqslant \rout$, 
\begin{subequations}
\begin{alignat}{2}
F &\sim l^{\,0}, \quad & 2 \leqslant l \leqslant 13, \label{F_lo} \\
F &\sim l^{\,\beta_1} e^{-\alpha_1 l}, \quad & 30 \leqslant l \leqslant 100. \label{F_hi}
\end{alignat}
\end{subequations}
We find that $\beta_1<0$, so $F$ decays slightly faster than an exponential at large $l$. The exact shape of $F$ is of secondary importance here. The key point is that there is no abrupt change in the scaling of $F$ with $l$ near the CMB.

\subsection{The magnetic time-variation spectrum $\FBt(l,r)$}

\FF{fig:FBtl_r}(a) plots the magnetic time-variation spectrum $\FBt$ at different $r$.
\begin{figure}
\centering
\includegraphics[width=0.45\textwidth]{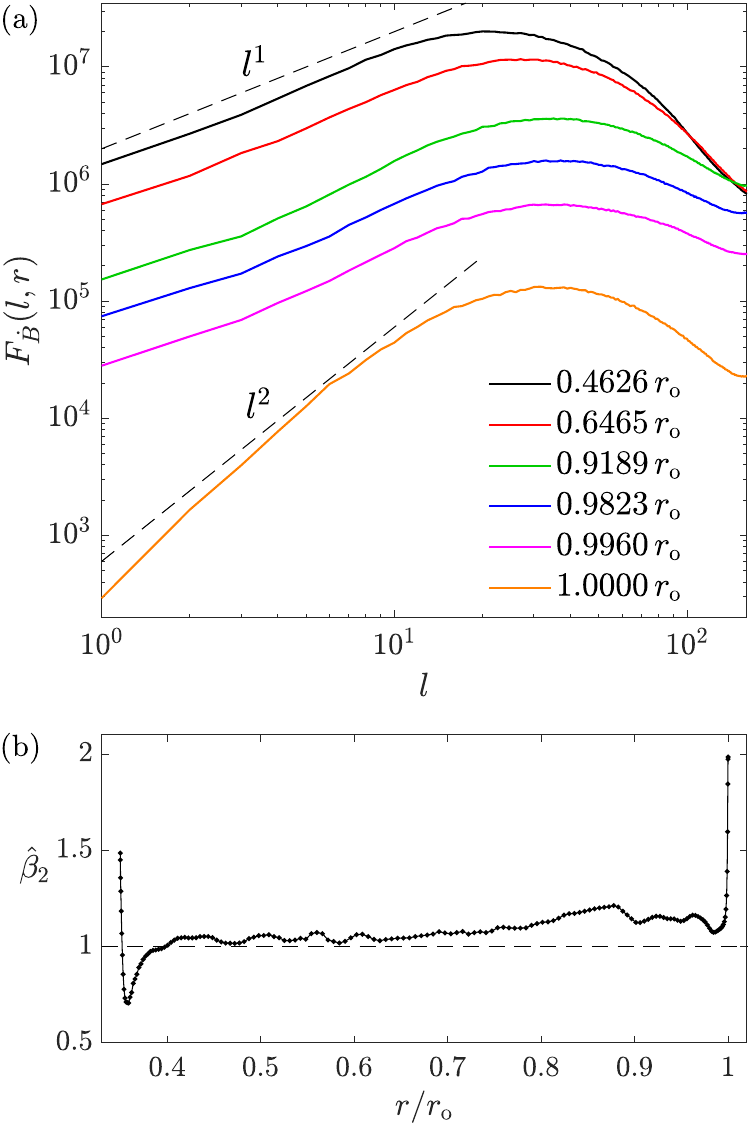}
\caption{(a) Time-averaged magnetic time-variation spectrum $\FBt(l,r)$, given by \Eq{FBtqst}, at different radius $r$ for run\,1. The scaling at small $l$ transitions from $\FBt \sim l$ in the interior to $\FBt \sim l^2$ at the CMB. The location $l = \lpeak$ of the peak varies weakly with $r$. (b) The exponent $\hat\beta_2$ in the power-law fit \Eq{hatbeta2} for $\FBt$ at small $l$ as a function of $r$.}
\label{fig:FBtl_r}
\end{figure}
For small $l$, specifically $2 \leqslant l \leqslant 13$, $\FBt$ increases with $l$ as a power-law for all $r$: $\FBt \sim l^{\,\hat\beta_2}$. The scaling exponent remains roughly constant at $\hat\beta_2=1$ in the interior and then transitions to $\hat\beta_2=2$ at the CMB. Thus for small $l$, $\FBt$ becomes steeper as $r$ increases. Furthermore this transition occurs within a thin boundary layer below the CMB, as shown in \Fig{fig:FBtl_r}(b) where we plot $\hat\beta_2$ versus $r$.

As $l$ increases, $\FBt$ eventually reaches its maximum value at some $l=\lpeak$ before decaying at the small scales, presumably due to magnetic diffusion. For run\,1, we find that $\lpeak$ generally increases weakly with $r$ from $\lpeak=25$ to $\lpeak = 35$ (for $r \gg \rin$). Similar to $F$, the decay at large $l$ can be approximated almost perfectly by an exponential modified by a power law for all $r$, as can be seen in \Fig{fig:F_hi}(b). Hence in summary,
\begin{subequations}
\begin{alignat}{2}
\qquad
\FBt &\sim l^{\,\hat\beta_2}, \quad & 2 \leqslant\,\ & l \leqslant 13, \label{hatbeta2} \\
\FBt &\sim l^{\,\beta_2} e^{-\alpha_2 l}, \quad & 30 \leqslant\,\ & l \leqslant 100, \label{FBt_hi}
\end{alignat}
\label{FBt_lo_hi}
\end{subequations}
where
\begin{equation}
\hat\beta_2 =
\begin{dcases}
1, & \quad \text{in the interior}, \\
2, & \quad \text{at the CMB}.
\end{dcases}
\label{hatb}
\end{equation}
We also note that $\beta_2>0$, so $\FBt$ decays slightly slower than an exponential at large $l$.

We expect $\lpeak$ to increase with $Rm$. So the Earth may have a larger $\lpeak$ if $Rm$ in the outer core is significantly higher than that in our simulations. It is not our goal to predict $\lpeak$ for the Earth. The proposal here is that the behaviour of $\FBt$ for the geodynamo at the large scales and the small scales, partitioned by some $\lpeak$, is similar to that described in \Eq{FBt_lo_hi} and \Eq{hatb}.

It is probably reasonable to assume $\lpeak$ also represents roughly where $F$ starts to drop off exponentially. Then by definition \Eq{taudef}, $\tau$ may have at most two different scaling regimes in $l$. The scale $\ltau$ that separates the two regimes is linked to $\lpeak$. Given that $\FBt$ has a rather flat peak, $\ltau$ most likely is less than but not too different from $\lpeak$. In the following, we adopt the usage that ``large scales" means $l < \ltau$ and ``small scales" means $l > \lpeak$.

\subsection{Connecting the scalings of $\FBt$ and $\tau$}
\label{sec:FBt_tau}

Equipped with this knowledge about $F$ and $\FBt$, we can understand the scaling of $\tau$ as follows. From the definition of $\tau$, together with \Eq{F_lo} and \Eq{hatb}, we have at the large scales:
\begin{equation}
\tau = \sqrt{\frac{F}{\FBt}} \sim \sqrt{\frac{l^0}{l^{\,\hat\beta_2}}}
\sim
\begin{dcases}
\sqrt{\frac{l^0}{l^1}} \sim l^{-1/2}, & \text{in the interior}, \\
\sqrt{\frac{l^0}{l^2}} \sim l^{-1},   & \text{at the CMB}.
\end{dcases}
\label{taul2}
\end{equation}
Hence the sharp change in the scaling of $\tau$ across the CMB boundary layer is mainly a consequence of the corresponding change in $\FBt$. At the small scales, we have seen that both $F_B$ and $\FBt$ are well approximated by an exponential modified by a power law for all $r$. So how does this lead to $\tau \sim l^{-1}$ at the small scales? \FF{fig:F_hi}(c) plots the differences $(\alpha_1 - \alpha_2)/2$ and $-(\beta_1 - \beta_2)/2$ between the exponents in \Eq{F_hi} and \Eq{FBt_hi}. It shows that $\alpha_1 \approx \alpha_2$ and $(\beta_1 - \beta_2)/2 \approx -1$, so the exponentials in \Eq{F_hi} and \Eq{FBt_hi} cancel each other resulting in $\tau \sim l^{-1}$.

\section{Balance of terms in the induction equation}

\Sect{sec:FBt_tau} establishes that the large-scale scaling of $\tau$ is predominantly linked to that of $\FBt$. Therefore in this section we focus on $\FBt$ and examine how its behaviour is controlled by the induction term and the diffusion term in the induction equation \Eq{indeq}.

\subsection{Spectrum of the induction term and spectrum of magnetic diffusion}
\label{sec:CHspec}

Let $\V G = \V u \times \V B$ and expand in terms of the vector spherical harmonics:
\begin{equation}
\V G = \sum_{l=0}^\infty \sum_{m=-l}^l  \big[ \qG(r,t) \VYlm + \sG(r,t) \VPsilm + \tG(r,t) \VPhilm \big].
\label{Adef}
\end{equation}
Then the induction term $\V C = \nabla\times(\V u\times \V B)$ can be written as a series of spherical harmonics modes:
\begin{multline}
\V C = \sumlm \bigg[\! -\frac{\sqrt{l(l+1)}\, \tG}r \VYlm - \frac{\big( r\tG \big)'}r \VPsilm \\
+ \frac{ \big( r\sG \big)' - \sqrt{l(l+1)}\, \qG}r \VPhilm \bigg],
\label{Cvsh}
\end{multline}
where $(\cdot)'$ denotes $r$-derivative. Now, in complete analogy to \Eq{Fdef}, we define the spectrum $F_C$ of $\V C$ by
\begin{equation}
\sum_{l=1}^\infty F_C(l,r,t) \equiv \frac{1}{4\pi} \oint |\V C(r,\theta,\phi,t)|^2 \dOmega.
\label{FCdef}
\end{equation}
Substituting \Eq{Cvsh} into \Eq{FCdef} gives $F_C$ in terms of the expansion coefficients $(\qG,\sG,\tG)$:
\begin{multline}
F_C(l,r,t) = \frac1{r^2} \frac{1}{(2l+1)}\sum_{m=0}^l \left[ l(l+1) \big| \tG \big|^2 + \big| \big( r\tG \big)' \big|^2 \right. \\
\left. + \big| \big( r\sG \big)' - \sqrt{l(l+1)}\qG\, \big|^2 \right] (4-3\delta_{m,0}).
\end{multline}
Similarly, for the diffusion term $\V D = \nabla^2 \V B = -\curl (\curl \V B)$, taking the double curl of the expansion \Eq{Bvsh} for $\V B$, we obtain
\begin{multline}
\V D = - \sumlm \bigg\{ \frac{ l(l+1) q_{lm} - \sqrt{l(l+1)} (r s_{lm})' }{r^2} \VYlm \\
+ \frac{ \sqrt{l(l+1)}q_{lm}' - (r s_{lm})'' }r \VPsilm
+ \bigg[ l(l+1)\frac{t_{lm}}{r^2} - \frac{(rt_{lm})''}r \bigg] \VPhilm \bigg\}.
\label{Hvsh}
\end{multline}
The spectrum of $\V D$ is then given in terms of the coefficients $(q_{lm},s_{lm},t_{lm})$ as:
\begin{multline}
F_D(l,r,t)
= \frac1{r^2} \frac{1}{(2l+1)} \sum_{m=0}^l \left[
\frac{l(l+1)}{r^2} \big|\! \sqrt{l(l+1)} q_{lm} - (r s_{lm})' \big|^2 \right. \\
\!\! \left.
 + \big| \sqrt{l(l+1)}q_{lm}' - (r s_{lm})''  \big|^2
 + \left| l(l+1)\frac{t_{lm}}r - (rt_{lm})'' \right|^2
\right] (4-3\delta_{m,0}). \!\!\!\!\!\!\!\!\!\!\!
\label{FHdef}
\end{multline}
Although $\V {\dot B} = \V C + \V D$, it is not true that $F_{\dot B} = F_C + F_D$, because cross-product terms arise when taking the squares for the spectra. However, when one of the three terms is small compared to the others, the spectra of the remaining two terms should be very close, and we exploit this below.

\begin{figure*}
\centering
\includegraphics[width=\textwidth]{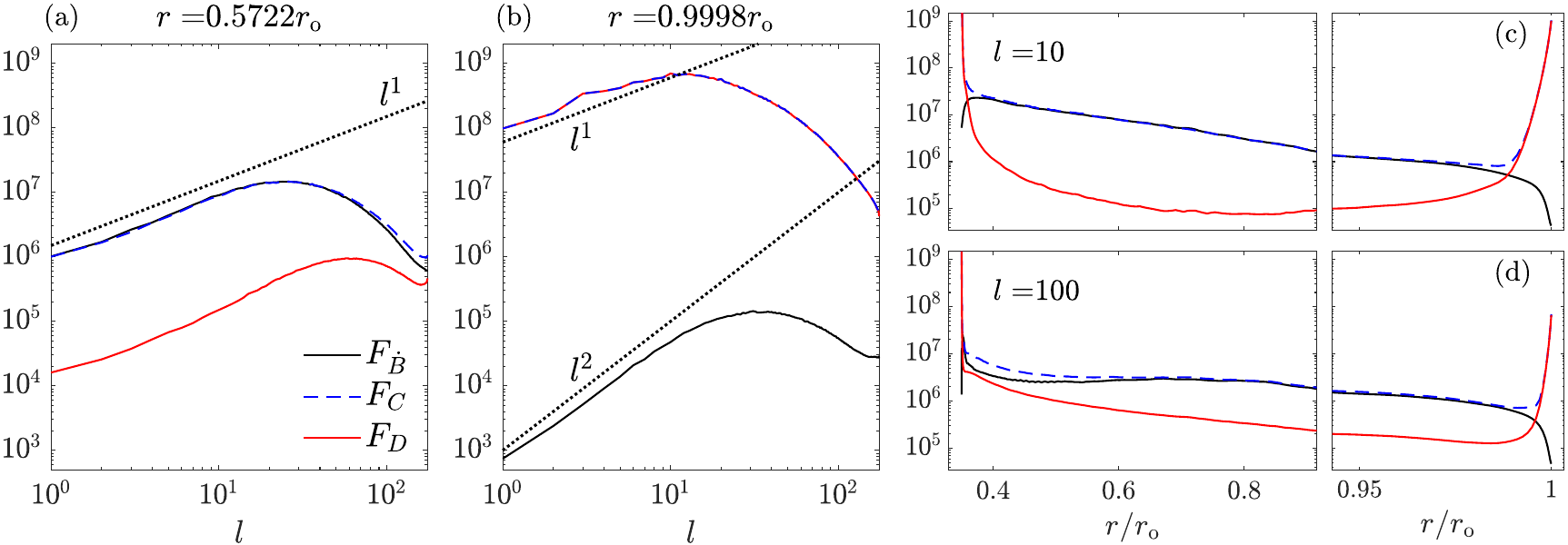}
\caption{The spectrum $F_C(l,r)$ of the induction term defined in \Eq{FCdef}, the spectrum $F_D(l,r)$ of the magnetic diffusion term given in \Eq{FHdef} and the magnetic time-variation spectrum $\FBt(l,r)$ of \Eq{FBtqst} for run\,1: (a) in the interior of the domain and (b) just below the CMB. (c) $F_C$, $F_D$ and $\FBt$ as a function of $r$ for the mode $l=10$. (d) Similar to (c) but for $l=100$.}
\label{BtCH0}
\end{figure*}

Let us first discuss the situation in the interior of the domain. This is demonstrated in \Fig{BtCH0}(a) which plots $F_C$ and $F_D$ together with $\FBt$ at $r=0.5722 \rout$.
It is clear that the diffusion term $\V D$ is negligible except for a small range of very large $l$. So to a very good approximation, $\FBt = F_C$ for most spatial scales. In particular for the large scales, \Fig{BtCH0}(a) shows that $F_C \sim l$ and thus from the arguments in \Eq{taul2}, the induction term alone is responsible for the large-scale scaling $\tau \sim l^{-0.5}$ in the interior. However, despite magnetic diffusion being negligible here, this scaling of $\tau$ cannot be predicted using the frozen-flux hypothesis.

Inside the CMB boundary layer, $F_C$ and $F_D$ increase sharply while $\FBt$ decreases as $r \to \rout$ from below. This is illustrated using the two modes $l=10$ and $l=100$ in \Figs{BtCH0}(c) and \ref{BtCH0}(d) respectively. The upshots are $F_C = F_D$ essentially and $\FBt \ll (F_C, F_D)$ for all $l$, as shown in \Fig{BtCH0}(b) for $r = 0.9998 \rout$. This suggests a dominant balance between the induction and the diffusion terms and the behaviour of $\dot{\V B}$ is a higher-order effect of this balance. Focusing on the large scales, we have the same scaling $F_C \sim l$ and $F_D \sim l$ as in the interior, as can be seen by comparing \Figs{BtCH0}(a) and \ref{BtCH0}(b). However the balance between $\V C$ and $\V D$ means these dominant scalings cancel each other, revealing a different scaling behaviour for $\FBt$, namely $\FBt \sim l^2$, which leads to $\tau \sim l^{-1}$ near the CMB. In anticipation of the stress-free results in \Sect{sec:sf}, we mention that the abrupt increase in $F_C$ and $F_D$ is related to the no-slip condition for the velocity.

In summary, we find that in the interior, magnetic diffusion is negligible in the spectra but the scaling $\tau \sim l^{-0.5}$ differs from the prediction based on the frozen-flux hypothesis. At the CMB, we indeed have $\tau \sim l^{-1}$, however, such scaling is achieved through boundary effects that involve the magnetic diffusion.

\subsection{The frozen-flux argument}

We now briefly review an argument that has been used to derive the scaling of $\tausv \sim l^{-1}$. We recall from \Eq{tausvdef} that in contrast to $\tau$, $\tausv$ is defined for $r \geqslant \rout$ only. Neglecting magnetic diffusion in the induction equation (the frozen-flux hypothesis), the radial magnetic field and the horizontal velocity $\uh = (0,u_\theta,u_\phi)$ at the CMB satisfy
\begin{equation}
\dot B_r = - \nabla_{\rm h} \cdot ( \uh B_r),
\label{frozen}
\end{equation}
where $\nabla_{\rm h} \cdot \V a = \nabla \cdot \V a - r^{-2} \partial (r^2 \unit r \cdot \V a)/\partial r$ for any vector $\V a$. Crudely, one may assume $\nabla_{\rm h} \sim \sqrt{l(l+1)} \sim l$ and $\uh \sim U$ where $U$ is some characteristic velocity scale independent of $l$. Then from \Eq{frozen}, it is argued that \cite[]{Holme06,Christensen12}
\begin{equation}
\tausv \sim \frac{B_r}{\dot B_r} \sim l^{-1}.
\label{tau_r}
\end{equation}
The keys to this ``frozen-flux argument" are magnetic diffusion is negligible and only the horizontal derivatives are important.

\subsection{Balance of terms in the spectral space}
\label{sec:specbal}

The frozen-flux argument heuristically connects the scaling of $\tau$ to the factor $\sqrt{l(l+1)}$ derived from the horizontal derivative. In this section, we consider the induction equation in the spectral space which allows us to carefully track the appearance of this factor. Combined with the finding that a very limited number of terms dominate the expansions of $\V C$ and $\V D$ in \Eq{Cvsh} and \Eq{Hvsh} respectively, we gain a deeper understanding of the results presented in \Sect{sec:CHspec}.

Substituting the expansions \Eq{Btvsh}, \Eq{Cvsh} and \Eq{Hvsh} into the induction equation \Eq{indeq} gives the following time evolution equations for the coefficients $q_{lm}$, $s_{lm}$ and $t_{lm}$:
\begin{subequations}
\begin{align}
\dot q_{lm} &= -\frac{\sqrt{l(l+1)}}r \tG -\frac{l(l+1)}{r^2} q_{lm} + \frac{\sqrt{l(l+1)}}r s_{lm}' + \frac{\sqrt{l(l+1)}}{r^2} s_{lm},
\label{qbal} \\
\dot s_{lm} &= -(\tG)' - \frac{\tG}r - \frac{\sqrt{l(l+1)}}r q_{lm}' + s_{lm}'' + \frac2r s_{lm}',
\label{sbal} \\
\dot t_{lm} &= (\sG)' + \frac{\sG}r - \frac{\sqrt{l(l+1)}}r \qG + t_{lm}'' + \frac2r t_{lm}' -\frac{l(l+1)}{r^2} t_{lm}.
\label{tbal}
\end{align}
\label{qsteq}%
\end{subequations}
Terms on the right-side of \Eq{qsteq} can be separated into two groups. Those that involve $\qG$, $\sG$ or $\tG$ originate from the induction term $\V C$ while terms involving $q_{lm}$, $s_{lm}$ or $t_{lm}$ come from the diffusion term $\V D$. A prefactor of $\sqrt{l(l+1)}$ is produced each time $\nabla_{\rm h}$ is applied in the physical space and again $(\cdot)'$ denotes the $r$-derivative. Note that the equations for $\dot q_{lm}$ and $\dot s_{lm}$ are not independent but are related in accordance with \Eq{qlmslm}.

To determine the dominant balance, we compare the spectra of the different terms in \Eq{qsteq}. The spectrum of any term is defined in an analogous manner to \Eq{Fqtdef}, e.g., the spectrum of the last term in \Eq{qbal} is
\begin{equation}
\frac{l(l+1)}{(2l+1)} \frac1{r^4} \sum_{m=0}^l \avg{ |s_{lm}|^2 }_t (4-3\delta_{m,0}).
\end{equation}
For the rest of this section, we focus exclusively on the large scales $l < \ltau$ where $\ltau \approx 13$ in our simulation. A key observation relevant to our discussion below is that terms that involve no derivative or only the radial derivative have fairly shallow spectra. On the other hand, the spectra of those terms involving the horizontal derivative have steep spectra dictated by the factor $l(l+1)$.

\begin{figure*}
\centering
\includegraphics[width=\textwidth]{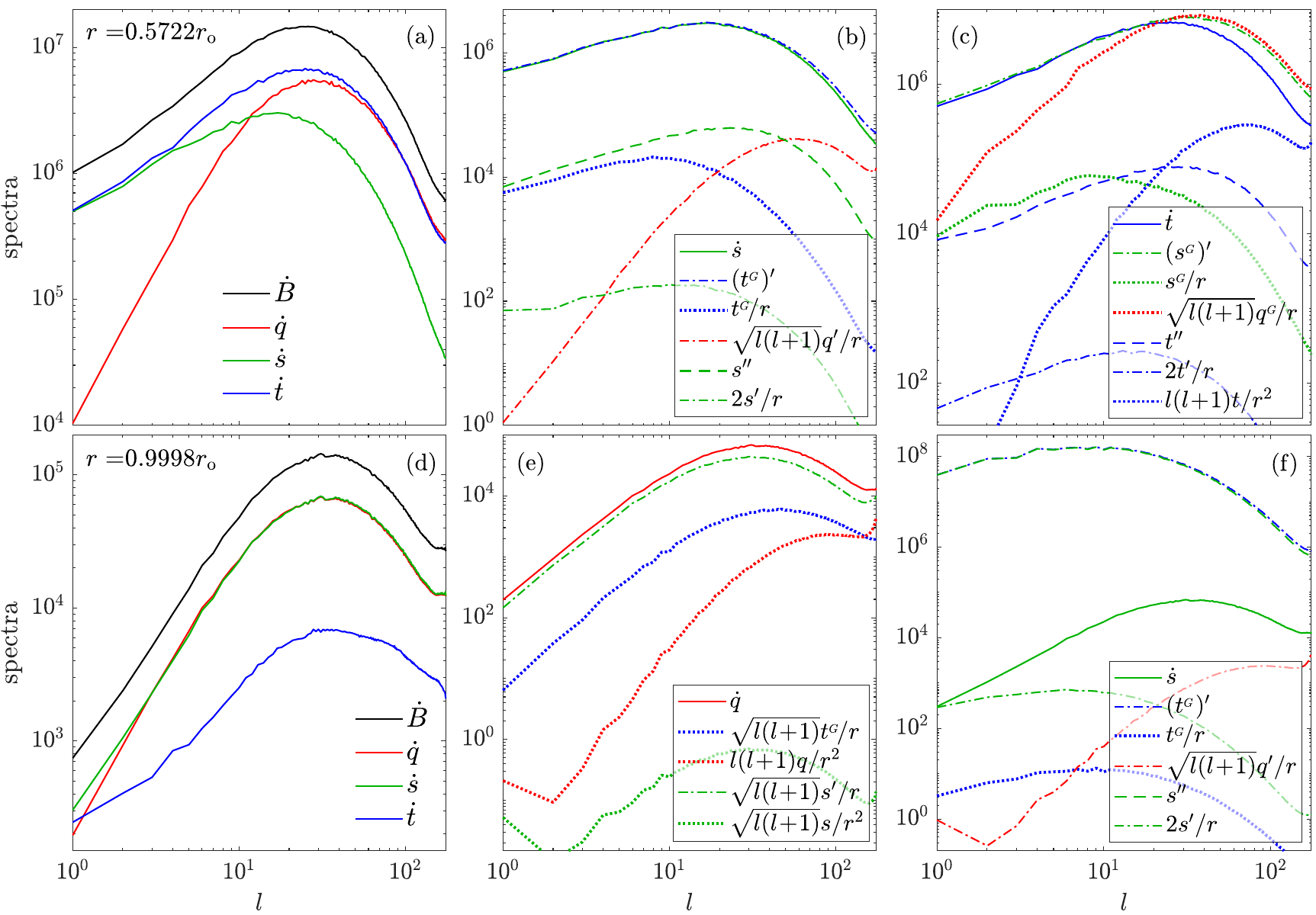}
\caption{For run\,1, the top row shows the spectral balance in the interior: (a) the magnetic time-variation spectrum $\FBt(l,r)$, given by \Eq{FBtqst}, and its three constituents $F_{\dot q}(l,r)$, $F_{\dot s}(l,r)$ and $F_{\dot t}(l,r)$ (see the text surrounding \Eq{Fqtdef} for definitions); (b) spectra of the individual terms in the time evolution equation \Eq{sbal} of $s_{lm}$, the subscripts $l,m$ are omitted in the legend for clarity; (c) spectra of the terms in the equation of motion \Eq{tbal} of $t_{lm}$. The bottom row shows the spectral balance near the CMB: (d) similar to (a) but for $r \approx \rout$, note that $F_{\dot t}$ eventually vanishes at $r=\rout$; (e) spectra of the terms in the time evolution equation \Eq{qbal} of $q_{lm}$; (f) similar to (b) but for $r \approx \rout$. It is clear that despite the large number of terms appearing in the equations, $\dot q_{lm}$, $\dot s_{lm}$ and $\dot t_{lm}$ are often dominated by only one or two terms.}
\label{Btqst}
\end{figure*}

We first consider the interior of the domain. With reference to the relation \Eq{FBtFqst}, \Fig{Btqst}(a) plots the spectra $F_{\dot q}$, $F_{\dot s}$, $F_{\dot t}$ and $\FBt$ at $r=0.5722 \rout$.
It shows that the steeper $F_{\dot q}$ is negligible and $\FBt$ is mainly composed of $F_{\dot s}$ and $F_{\dot t}$. Moreover, both $F_{\dot s}$ and $F_{\dot t}$ scales like $l^1$, i.e. the same scaling of $\FBt$ which leads to $\tau \sim l^{-0.5}$. \FF{Btqst}(b) plots $F_{\dot s}$ together with the spectra of the terms on the right-side of \Eq{sbal}. It clearly shows that, among the many terms, $\dot s_{lm}$ is dominated by a single one, namely $(\tG)'$. Similarly \Fig{Btqst}(c) shows $\dot t_{lm}$ is dominated by $(\sG)'$ in \Eq{tbal}. In summary, for the large scales in the interior:
\begin{subequations}
\begin{align}
\FBt &\approx F_{\dot s} + F_{\dot t} \sim l, 
\label{lsbalin1} \\
\dot q_{lm} &\approx -\frac{\sqrt{l(l+1)}}r \tG,
\label{lsbalinq} \\
\dot s_{lm} &\approx -(\tG)',
\label{lsbalin2} \\
\dot t_{lm} &\approx (\sG)'.
\label{lsbalin3}
\end{align}
\label{lsbalin}%
\end{subequations}
The balance for $\dot q_{lm}$ in the interior is not relevant for the present discussion and is not plotted in \Fig{Btqst}, but it is included in \Eq{lsbalin} for later discussion.

From \Eq{lsbalin}, we infer two properties of the large-scale magnetic field in the interior. Firstly, the absence of $F_{\dot q}$ in \Eq{lsbalin1} implies the main contribution to $\dot{\V B}$ comes from $\dot B_\theta$ and $\dot B_\phi$. This echoes the discussion near the end of \Sect{sec:tauPT}. Secondly, \Eq{lsbalin2} and \Eq{lsbalin3} tell us that the scaling of $\FBt$ is related to the radial derivatives in the induction term $\V C$ rather than the horizontal derivatives. This is the reason why the prediction of $\tau \sim l^{-1}$ by the frozen-flux argument is not valid even though magnetic diffusion is negligible here.

It is not useful to analyse \Eq{qsteq} near the bottom of the CMB boundary layer where the transition in scaling starts as a large number of terms are involved. However, the situation becomes simple again at the CMB and in the vicinity just below it. This is illustrated in \Fig{Btqst}(d)--(f) for $r=0.9998 \rout$ and summarised below (for the large scales):
\begin{subequations}
\begin{align}
\FBt &\approx F_{\dot q} + F_{\dot s} \sim l^2, \label{lsbalout2} \\
\dot q_{lm} &\approx \frac{\sqrt{l(l+1)}}r s_{lm}', \label{lsbalout3}\\\
\dot s_{lm} &\approx -(\tG)' + s_{lm}'', \label{lsbalout4} \\
\dot t_{lm} &\to 0,
\end{align}
\label{lsbalout}%
\end{subequations}
with some additional contribution to $\dot s_{lm}$ at very small $l$ coming from $2s_{lm}'/r$ (we have verified that the spectrum of the sum $-(\tG)' + s_{lm}''$ is indeed close to $F_{\dot s}$).

The absence of $F_{\dot t}$ in \Eq{lsbalout2} reflects the discussion in \Sect{sec:tauPT}: $\dot{\V B} \approx \dot{\V B}_{\rm Pol}$ near the CMB with $\BPol$ related to $q_{lm}$ and $s_{lm}$ by \Eq{BPqslm}. We also expect $F_{\dot q}$ and $F_{\dot s}$ to be approximately equal because of \Eq{slm_qlm} required by the magnetic boundary condition. Indeed, \Fig{Btqst}(d) shows that $F_{\dot q} \approx F_{\dot s} \sim l^2$ and the two spectra contribute equally to the sum in \Eq{lsbalout2}. We also know from \Sect{sec:CHspec} that magnetic diffusion is involved in the scaling $\FBt \sim l^2$ at the CMB. Further insights are provided by \Eq{lsbalout}. It can be proved that the no-slip condition implies $\tG=0$ at the CMB. So $\dot q_{lm}$ must be controlled by the diffusion terms on the right-side of \Eq{qbal}. It turns out the dominant balance is \Eq{lsbalout3}. Since \Fig{Btqst}(f) shows that the spectrum of $s_{lm}'$ is quite flat, it immediately follows from \Eq{lsbalout3} that $F_{\dot q} \sim l^2$. For the balance of $\dot s_{lm}$, it can be seen in \Fig{Btqst}(f) that the two terms on the right-side of \Eq{lsbalout4} have shallow spectra. However, the almost exact balance of these two terms results in the steeper scaling of $F_{\dot s} \sim l^2$. This is the origin of the similar cancellation between $F_C$ and $F_D$ depicted in \Fig{BtCH0}(b) and described in details near the end of \Sect{sec:CHspec}. We also note that in \Eq{lsbalout4}, $s_{lm}''$ is a diffusion term and both $(\tG)'$ and $s_{lm}''$ involve the radial derivative rather than the horizontal derivative, so none of these is compatible with the frozen-flux argument.

\subsection{Spectral balance for the radial component}
\label{sec:spec_r}

\begin{figure}
\centering
\includegraphics[width=0.45\textwidth]{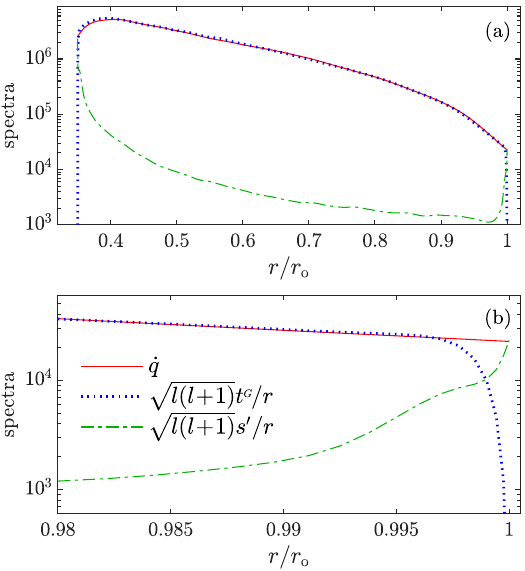}
\caption{(a) The spectra of the three terms in \Eq{qdotbal} for run\,1 as a function of radius $r$ for the $l=10$ mode (the subscripts $l,m$ are omitted in the legend). This shows the transition in the balance of $\dot q_{lm}$ from induction-dominant in the interior to diffusion-dominant near the CMB. The situation is similar for large $l$ (b) Same data as in (a) but zooming into the boundary layer to show the transition in details. We have verified that the spectra of the sum $\sqrt{l(l+1)}(-\tG + s_{lm}')/r$ (not shown) virtually overlaps $F_{\dot q}$ (red solid curve) for all $r$, including inside the boundary layer.}
\label{lsqdotbal}
\end{figure}

The results presented in \Sect{sec:specbal} reaffirm that the frozen-flux argument does not predict the scaling of $\tau$. But how about $\tau_r$? After all, strictly speaking, the frozen-flux argument is based on $B_r$. We investigate this by first reporting that $F_q \sim l^0$ at the large scales for all $r$ (not shown), so the scaling of $\tau_r$ is determined by $F_{\dot q}$. The dominant balance of $\dot q_{lm}$ is given by \Eq{lsbalinq} for the interior and \Eq{lsbalout3} near the CMB. Most interestingly, we find that by combining \Eq{lsbalinq} and \Eq{lsbalout3}, the following simple expression works very well at all $r$:
\begin{equation}
\dot q_{lm} \approx -\frac{\sqrt{l(l+1)}}r \tG + \frac{\sqrt{l(l+1)}}r s_{lm}'.
\label{qdotbal}
\end{equation}
\FFs{Btqst}(b) and \ref{Btqst}(f) show that, for different $r$, the spectra of both $\tG$ and $s'_{lm}$ vary weakly with $l$ at the large scales, so \Eq{qdotbal} gives $F_{\dot q} \sim l^2$ for all $r$. This implies the scaling $\tau_r \sim l^{-1}$ is maintained as $r \to \rout$ from below although the origin of this scaling stealthily transitions from induction in the interior to diffusion near the CMB. The variation of the different terms in \Eq{qdotbal} with $r$ is demonstrated in \Fig{lsqdotbal} using the $l=10$ mode.

To answer the question posed at the beginning of this subsection, we find that the frozen-flux argument correctly predict the scaling of $\tau_r$ in the interior which arises from the horizontal derivative of the induction term. This is not surprising since the balance \Eq{lsbalinq} is essentially the spectral version of \Eq{frozen}. The argument fails when $\tG \to 0$ near the CMB as required by the no-slip condition. We shall see in the next section that the situation is different with a stress-free condition at the CMB.

\section{Stress-free boundary condition}
\label{sec:sf}

\begin{figure*}
\centering
\includegraphics[width=\textwidth]{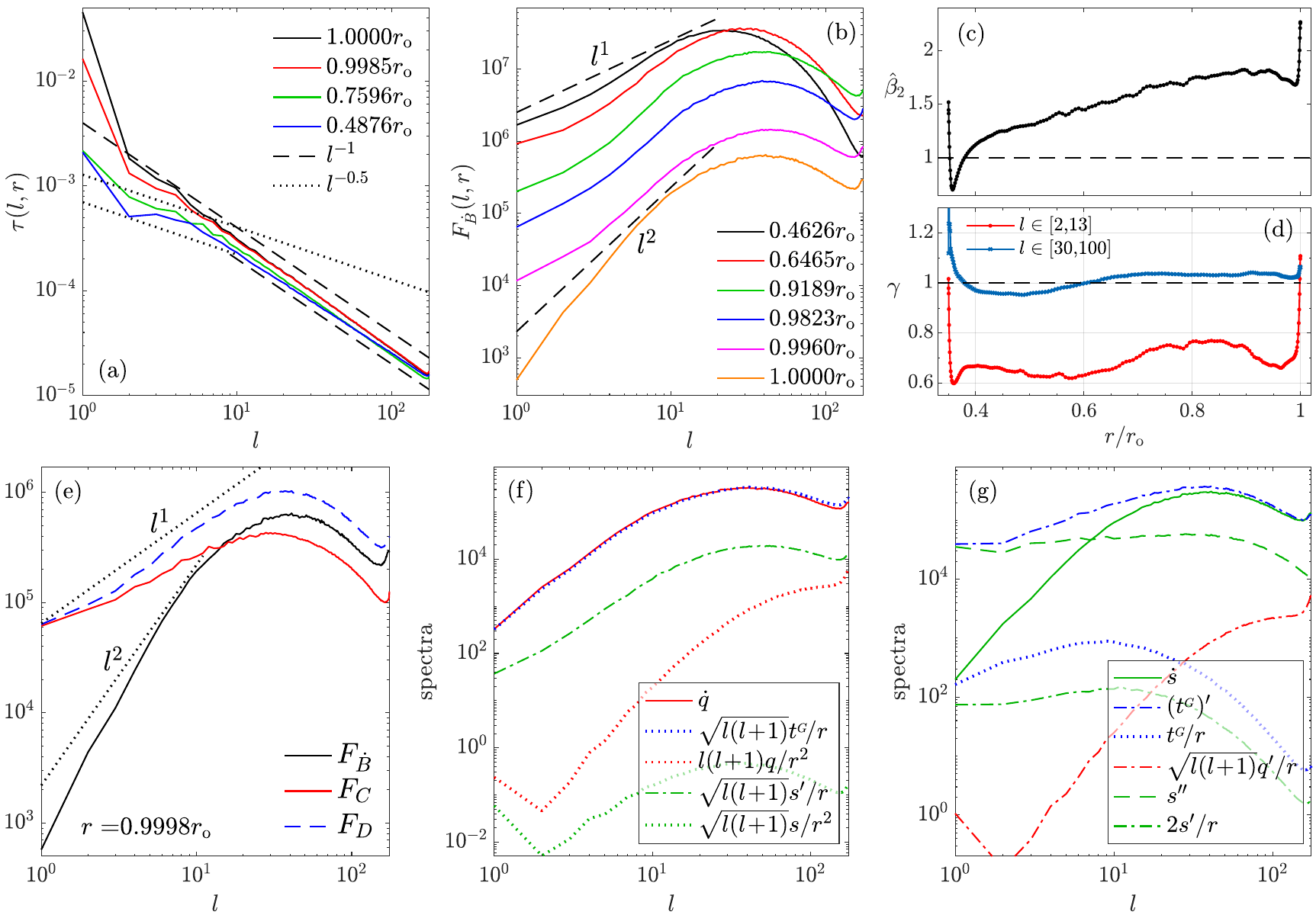}
\caption{Summary of results for run\,2 which has a stress-free condition at the CMB. (a) Time-averaged magnetic time-scale spectrum $\tau(l,r)$ at different radius $r$ (b) Time-averaged magnetic time-variation spectrum $\FBt(l,r)$ at different $r$. $\FBt$ is slightly steeper than $l^2$ at the CMB. (c) The exponent $\hat\beta_2$ in the power-law fit $\FBt \sim l^{\hat\beta_2}$ for the large scales as a function of $r$. This shows $\hat\beta_2$ varies considerably in the interior. (d) The scaling exponent $\gamma$ in the power-law fit \Eq{taupwrlaw} to $\tau$ as a function of $r$. The least-square fit is performed separately for the large scales (red) and the small scales (blue). (e) The spectrum $F_C$ of the induction term, the spectrum $F_D$ of the diffusion term and $\FBt$ near the CMB. (f) Spectra of the terms in the time evolution equation \Eq{qbal} of $q_{lm}$ for $r=0.9998\rout$. The subscripts $l,m$ are omitted in the legend. (g) Similar to (f) but for the time evolution equation \Eq{sbal} of $s_{lm}$.}
\label{fig:sf}
\end{figure*}

We have learned that boundary conditions play a role in the transition of $\tau$ and $\FBt$ near the CMB. In order to distinguish between the influence of the magnetic boundary condition and that of the velocity boundary condition, we study the geodynamo model \Eq{ndmodel} using the stress-free condition for the velocity at the CMB in run\,2. Other boundary conditions and all simulation parameters are the same as in run\,1 which uses the no-slip condition (see Table~\ref{para_tbl}).

With the stress-free condition, the three spectra $F$, $\FBt$ and $\tau$ behave qualitatively the same as in run\,1, excepting some minor differences for $\tau$ and $\FBt$, which are plotted in \Figs{fig:sf}(a) and \ref{fig:sf}(b) respectively. These are to be compared with \Figs{fig:tau}(a) and \ref{fig:FBtl_r}(a) for run\,1 with the no-slip condition. In run\,1, the scaling of $\FBt$ at $l < \lpeak$ remains roughly the same throughout the interior. Here, \Fig{fig:sf}(c) shows that $\hat\beta_2 \approx 1$ near the inner boundary and increases gradually to $\hat\beta_2 \approx 1.75$ just below the CMB boundary layer. As a result, $\tau$ at the large scales is steeper than $l^{-0.5}$ by a fair amount with $0.6 \lesssim \gamma(r) \lesssim 0.8$ as shown in \Fig{fig:sf}(d).

Just as in run\,1, there is a sharp increase of steepness in $\tau$ and $\FBt$ when $r \to \rout$ from below, as illustrated by $\hat\beta_2(r)$ in \Fig{fig:sf}(c) and $\gamma(r)$ in \Fig{fig:sf}(d). This shows that the no-slip condition is not crucial for this transition to occur. At the CMB, we have $\hat\beta_2 \approx 2.26$ and $\gamma \approx 1.11$ which are not different in a significant manner from the values in run\,1. This means that, disappointingly, we cannot deduce the velocity boundary condition at the CMB by observing $\tausv$ at the planetary surface. We also find that the spectra $\tauPol$, $\tauTor$ and $\tau_r$ all follow the same pattern described in \Sect{sec:tauPT} and \Fig{fig:tauPT}. In other words, their scalings are also insensitive to the velocity boundary condition.

Regarding the balance of terms in the induction equation, the situation in the interior is almost exactly the same as in run\,1 shown in \Fig{BtCH0}(a), i.e., $\FBt \approx F_C$ with $F_D$ negligible. Similarly, \Figs{Btqst}(a)--(c) and the relations in \Eq{lsbalin} describe the spectral balance in the interior equally well for the stress-free case. Near the CMB, the magnetic boundary condition necessitates $F_{\dot q} \approx F_{\dot s}$ and we find that $\FBt \approx F_{\dot q} + F_{\dot s} \sim l^2$, again the same as in run\,1. We shall see the main difference between the stress-free and no-slip boundary conditions is finally unveiled when we investigate the balance of terms near the CMB.

With the stress-free boundary condition, \Fig{fig:sf}(e) shows that near the CMB, $F_C$ and $F_D$ are no longer many orders of magnitude larger than $\FBt$ and neither do they overlap, in contrast to \Fig{BtCH0}(b) for run\,1. Therefore the sharp rise in $F_C$ and $F_D$ seen in \Fig{BtCH0} is caused by the no-slip condition at the CMB. Closely related to this is the balance in the spectral space for $\dot s_{lm}$. We find that the dominant balance \Eq{lsbalout4} at the large scales is also valid in the stress-free case. However, \Fig{fig:sf}(g) shows that the spectra of $(\tG)'$ and $s_{lm}''$ do not become much larger than $F_{\dot s}$ near the CMB, unlike in \Fig{Btqst}(f) for run\,1.

We now turn to the balance for $\dot q_{lm}$. An important consequence of employing the stress-free condition at the CMB is that $\tG \neq 0$ at $r = \rout$. We then find that the relation \Eq{lsbalinq} is now valid for all $r$, including near the CMB as shown in \Fig{fig:sf}(f). The switching to a diffusion term near the CMB depicted in \Fig{lsqdotbal} for run\,1 does not occur here. Since \Eq{lsbalinq} is effectively the spectral form of the frozen-flux hypothesis and it leads to $\tau_r \sim l^{-1}$, we conclude that in the stress-free case, the frozen-flux argument correctly predicts the scaling of $\tau_r$ at the large scales everywhere (away from the inner boundary). However, for $\tau$ near the CMB, since both $F_{\dot q}$ and $F_{\dot s}$ contribute and the dominant balance of $\dot s_{lm}$ involves the diffusion term $s_{lm}''$, the frozen-flux hypothesis is violated.

\section{Effects of varying $E\lowercase{k}$ and $R\lowercase{m}$}
\label{EkRm}

\begin{figure*}
\centering
\includegraphics[width=0.98\textwidth]{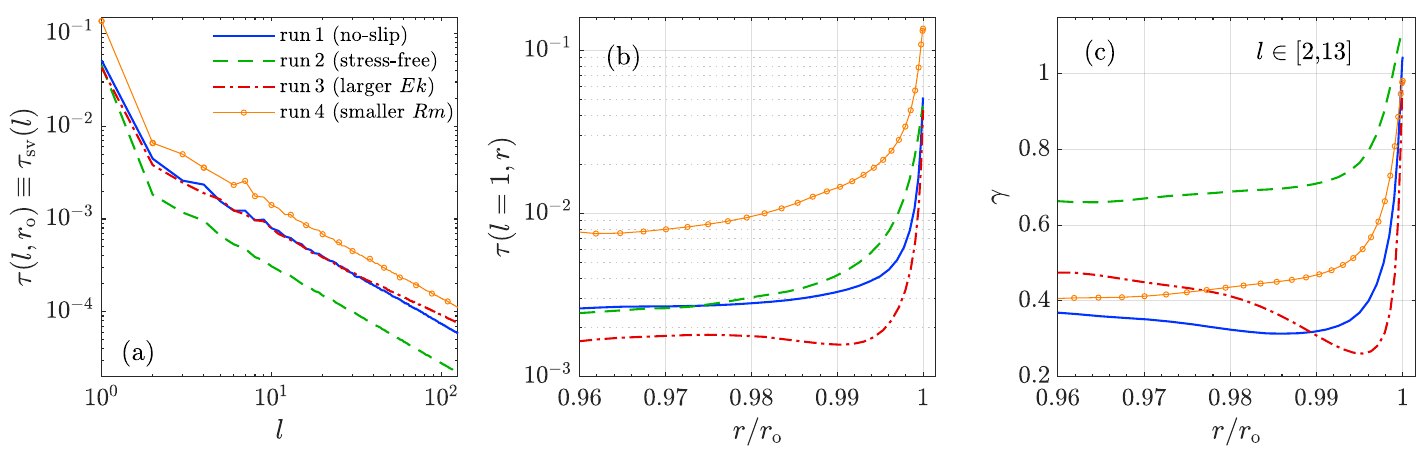}
\caption{Comparison of the time-variation spectrum $\tau(l,r)$ for the four runs listed in Table\,\ref{para_tbl}: (a) $\tau(l,r)$ versus $l$ at the CMB $r=\rout$; (b) $\tau(l,r)$ as a function of $r$ for the dipole $l=1$ near the CMB; (c) the scaling exponent $\gamma$ in the power-law fit \Eq{taupwrlaw} to the large scales of $\tau$ as a function of $r$ near the CMB. The range of $l$ used in the fit is shown in the figure.}
\label{fig:cmb_bl}
\end{figure*}

In this section, we consider two simulations with the no-slip condition that have simulation parameters different from those in run\,1. Referring to Table~\ref{para_tbl}, run\,3 has a larger $Ek$ than run\,1 and we also adjust $Ra$ so that $Rm$ is approximately the same as in run\,1. Run\,4 uses a smaller $Ra$, resulting in a $Rm$ that is about half of that in run\,1.
 
The scalings of the spectra $F$, $\FBt$ and $\tau$ in both run\,3 and run\,4, as well as how these scalings vary with $r$ and $l$, are essentially the same as what we have found in run\,1. We also find that, as expected, run\,4 has a smaller $\lpeak$ than run\,1, indicating that $\lpeak$ increases with $Rm$. \FF{fig:cmb_bl}(a) plots $\tausv(l)$ from different runs. Comparing run\,1 and run\,3, it appears that the effect of $Ek$ on $\tausv$ is small. This is also true for $\tau(l,r)$ in the interior, especially at large $l$. Comparing run\,4 to run\,1 (or run\,3), \Fig{fig:cmb_bl}(a) suggests that $\V B$ varies on increasingly shorter time scales as $Rm$ increases. However we note that as $r$ decreases, $\tau$ from run\,2, which has the highest $Rm$ but also different boundary conditions, approaches those of run1 and run 3 from below. It may be because the local $Rm$ in run\,2 is smaller in the interior than near the CMB.
 
We now focus on the transition layer below the CMB in which the scaling of $\tau$ changes. \FFs{fig:cmb_bl}(b) and \ref{fig:cmb_bl}(c) show, respectively, $\tau(l=1,r)$ and $\gamma$ for the large scales as a function of $r$ just below the CMB. Comparing run\,1 and run\,4 shows that, unsurprisingly, the thickness of the transition layer increases as $Rm$ decreases. By contrast, the effect of $Ek$ is more curious. Previous sections establish that the scaling $\tau \sim l^{-\gamma}$ undergoes a transition regardless of whether an Ekman layer exists near the CMB. However, the presence of an Ekman layer does change the characteristics of the transition. Comparing run\,3 ($Ek$\,=\,$1 \times 10^{-4}$), run\,1 ($Ek$\,=\,$2.5 \times 10^{-5}$) and run\,2 (stress-free) in \Figs{fig:cmb_bl}(b) or \ref{fig:cmb_bl}(c) indicates that as $Ek$ decreases, the thickness of the transition layer increases towards that in run\,2. It is not clear in run\,2 whether this transition layer associated with $\tau$ coincides with the magnetic diffusive layer \cite[]{TerraNova20}. Further investigation is needed to better understand the interaction between the magnetic field and the velocity inside the CMB boundary layer.

\section{Discussion}

In all of our simulations, either with a no-slip condition or a stress-free condition at the outer boundary, $\tausv(l) \equiv \tau(l,\rout) \sim l^{-1}$ is found at the CMB. So the scaling of $\tausv$ provides no hint about the boundary condition at the CMB. And yet this scaling actually has a very different origin in these two types of model. This is most clearly demonstrated in $\tau_r$ (which equals $\tausv$ at $r=\rout$). In the no-slip case, the scaling $\tau_r \sim l^{-1}$ at the CMB comes from the diffusion term $\sqrt{l(l+1)} s_{lm}'/r$ while in the stress-free case, the very same scaling originates from the induction term $\sqrt{l(l+1)} \tG/r$. It is sometimes argued that $\tausv \sim l^{-1}$ implies the secular variation is driven by the induction term whereas $\tausv \sim l^{-2}$ indicates it is due to magnetic diffusion \cite[]{Holme06, Christensen12, Sharan22}. Here we see that this interpretation is an oversimplification.

Regardless of the velocity boundary condition, the scaling of $\tau$ at the large scales near the CMB is always different from that in the interior, suggesting the large-scale magnetic field varies on different time scales in the two regions. Hence once again, details about the magnetohydrodynamics in the interior are hidden from $\tausv$ at the surface. In hindsight, this is not too surprising as the condition of $\V B$ being potential at $r = \rout$ combined with $\nabla \cdot \V B = 0$ tie $\dot B_r$, $\dot B_\theta$ and $\dot B_\phi$ together at and above the CMB, preventing the genuine time variation of $(B_\theta,B_\phi)$ in the interior from being observed above the CMB. Maybe the real surprise here is why only the large scales are affected, with $\tau \sim l^{-1}$ found at the small scales for all $r$.

\subsection{Inferring $\tau$ at the small scales in the interior}

While our focus is mainly on the large scales, we now briefly discuss the situation at the small scales. Difficulties in the measurement of the internal magnetic field $\V B$ at the small scales due to the presence of the lithospheric field means $\tausv(l)$ is not available from observation for $l \gtrsim 13$. A result in our simulations and in previous numerical studies \cite[]{Christensen12} is that $\tausv$ exhibits the same scaling for all $l$. Thus we suggest that the scaling at the large scales derived from observation can be extrapolated to the small scales. Another interesting point of our investigation is that $\tau(l,r)$ varies only weakly with $r$ for large $l$. This is illustrated in \Fig{fig:tau}(b) for run\,1. The dependence on $r$ is even weaker for the stress-free case of run\,2, as can be inferred from \Fig{fig:sf}(a). The significance of these findings is that even if $F$ and $\FBt$ are not available at the small scales, it is possible to estimate the time scale of a small-scale mode $\lhi$ inside the outer core using large-scale properties observed at the surface. For a radius $r_1$ in the interior, the estimation goes as follows:
\begin{equation}
\tau(\lhi,r_1) \approx \tau(\lhi,\rout) \equiv \tausv(\lhi) \approx \tau_*(\rout) \lhi^{-\gamma(\rout)},
\label{taulhi}
\end{equation}
where $\tau_*(\rout)$ and $\gamma(\rout)$ are obtained from a power-law fit to the accessible large-scale modes of $\tausv$.

Regarding the scaling law at the small scales, we find $\gamma \approx 1$ for all $r \gg \rin$ in our simulations. This is irrespective of the velocity boundary condition used at the CMB, as shown in \Fig{fig:gamma}(a) and \Fig{fig:sf}(d). A physical meaning can be attached to the scaling $\tau \sim l^{-1}$ by invoking Jeans' rule \cite[]{Jeans23} which relates the spatial scale $\lambda$ of a feature to the spherical harmonics degree $l$ by
\begin{equation}
\lambda = \frac{2\pi \rE}{\sqrt{l(l+1)}}.
\end{equation}
Then $\gamma=1$ means $\tau$ is directly proportional to $\lambda$ because
\begin{equation}
\tau = \frac{\tau_*}l = \tau_* \bigg[ \frac1{\sqrt{l(l+1)}} + O(l^{-2}) \bigg] \approx \frac{\tau_*}{2\pi\rE} \lambda.
\label{taulambda}
\end{equation}
If we interpret $\tau$ as a turn-over time over which a ``magnetic eddy’’ of size $\lambda$ completely reorganises \cite[]{Stacey92}, then \Eq{taulambda} implies doubling the size of a small-scale magnetic eddy allows it to live twice as long.

\subsection{Inferring $\tau$ at the large scales in the interior}

The scaling of the secular variation spectrum $\Rsv(l) \equiv \FBt(l,\rout) \sim l^2$ at the large scales found in our simulations is consistent with most of the results based on satellite observations, for example, see \cite{Gillet10}, \cite{Holme11} and more recently, \cite{Finlay20} which extends the observations of $\Rsv$ up to $l \approx 17$, thanks to the fact that $\V{\dot B}$ is less polluted by the static lithospheric field than $\V B$ itself. Beyond $l \approx 17$, it is currently unclear how $\Rsv$ behaves. Hypothetically, if the shape of $\Rsv$ is similar to our simulation results in \Fig{fig:FBtl_r}(a) or \Fig{fig:sf}(b) and $\lpeak$ can be determined, we can make an order-of-magnitude estimate on the time scale of a large-scale mode $\llow < l_\eta$ at some radius $r_1$ in the interior. We assume $\tau = \tau_* l^{-1/2}$ for $l < l_\tau$ in the interior. Making the approximation $l_\tau = l_\eta$, we have the estimate
\begin{equation}
\tau(\llow,r_1) = \tau(l_\eta,r_1) \bigg( \frac{\llow}{l_\eta} \bigg)^{-\frac12}
= \tau_*(\rout) \cdot (l_\eta \llow)^{-\frac12},
\label{taul1r1}
\end{equation}
where in the last step we have used \Eq{taulhi} with $\lhi = l_\eta$ to deduce $\tau(l_\eta,r_1)$ and we also set $\gamma(\rout)=1$ for simplicity. Since our argument neglects the variation of $\tau$ with $r$ in the interior, $r_1$ does not appear on the right-side of \Eq{taulhi} and \Eq{taul1r1}. For run\,1, $\tau_*(\rout) = 9 \times 10^{-3}$ and $l_\eta = 31$. At $r_1 = 0.7596 \rout$, for $\llow=1$ and $\llow=5$, the values of $\tau$ from the simulation data are $2.28 \times 10^{-3}$ and $6.15 \times 10^{-4}$ respectively. Applying \Eq{taul1r1}, we obtain the estimates of $\tau = 1.62 \times 10^{-3}$ for $\llow=1$ and $\tau = 7.23 \times 10^{-4}$ for $\llow=5$. So if the value of $\lpeak$ becomes available in the future---admittedly a big if---$\tausv$ may give us some information on the large-scale dynamics inside the outer core.

\begin{acknowledgments}

The authors are supported by the Science and Technology Facilities Council (STFC), `A Consolidated Grant in Astrophysical Fluids' (grant numbers ST/K000853/1 and ST/S00047X/1). This work used the DiRAC Complexity system, operated by the University of Leicester IT Services, which forms part of the STFC DiRAC HPC Facility (www.dirac.ac.uk). This equipment is funded by BIS National E-Infrastructure capital grant ST/K000373/1 and  STFC DiRAC Operations grant ST/K0003259/1. DiRAC is part of the National E-Infrastructure.

The authors would like to thank the Isaac Newton Institute for Mathematical Sciences, Cambridge, for support and hospitality during the programme DYT2 where part of this paper was written. DYT2 was supported by EPSRC grant EP/R014604/1.\\

\end{acknowledgments}

\noindent
{\bf DATA AVAILABILITY}\\[0.2cm]
The numerical code used for the geodynamo simulations reported in this paper is at https://github.com/Leeds-Spherical-Dynamo. For access to the Github repository, please contact the authors.

\bibliography{geo_tau}

\appendix

\section{Vector spherical harmonics}
\label{vsh}

Using the Schmidt semi-normalised associated Legendre polynomials $P_l^m$, we define the spherical harmonics as follow,
\begin{equation}
\Ylm(\theta,\phi) = P_l^{|m|}(\cos\theta) e^{im\phi}.
\label{Ylmdef}
\end{equation}
Then closely following \cite{Barrera85}, we define the following set of vector spherical harmonics:
\begin{subequations}
\begin{align}
\VYlm(\theta,\phi) &= \Ylm \V{\hat r}, \label{VYlmdef} \\
\VPsilm(\theta,\phi) &= \frac{1}{\sqrt{l(l+1)}}\, r \nabla \Ylm, \\
\VPhilm(\theta,\phi) &= \V{\hat r} \times \VPsilm, \label{VPhilmdef}
\end{align}
\end{subequations}
which form an orthogonal basis for all square-integrable vector fields on the unit sphere. Note that $\VPsilm$ and $\VPhilm$ are defined for $l>0$. The semi-normalisation condition is
\begin{gather}
\oint \VYlm \cdot (\V Y_{l'}^{m'})^* \dOmega
= \frac{4\pi}{2l+1}(2-\delta_{m,0})\delta_{ll'}\delta_{mm'},
\end{gather}
with similar expressions for $\VPsilm$ and $\VPhilm$.

The magnetic field $\V B$ can be expanded in terms of $\{ \VYlm, \VPsilm, \VPhilm \}$ as in \Eq{Bvsh}, which is repeated below:
\begin{equation}
\V B = \sumlm \left( q_{lm} \VYlm + s_{lm} \VPsilm + t_{lm} \VPhilm \right).
\tag{\ref{Bvsh}}
\end{equation}
In terms of the expansion coefficients $(q_{lm},s_{lm},t_{lm})$, the components of $\V B$ are:
\begin{subequations}
\begin{align}
B_r &= \sumlm q_{lm} \Ylm, \label{Brqlm} \\
B_\theta &= \sumlm \frac1{\sqrt{l(l+1)}}
\left( s_{lm} \pp{\Ylm}{\theta} - \frac{t_{lm}}{\sin\theta} \pp{\Ylm}{\phi} \right), \\
B_\phi &= \sumlm \frac1{\sqrt{l(l+1)}}
\left( t_{lm} \pp{\Ylm}{\theta} + \frac{s_{lm}}{\sin\theta} \pp{\Ylm}{\phi} \right).
\end{align}
\label{Brtpqst}%
\end{subequations}
The magnetic field can also be written as the sum of a poloidal part and a toroidal part as in \Eq{BPT}, and the scalar potentials $\Pol$ and $\Tor$ can be expanded in terms $\Ylm$:
\begin{subequations}
\begin{align}
\Pol(r,\theta,\phi,t) &= \sumlm \widetilde \Pol_{lm}(r,t) \Ylm(\theta,\phi), \label{PTlm1} \\
\Tor(r,\theta,\phi,t) &= \sumlm \widetilde \Tor_{lm}(r,t) \Ylm(\theta,\phi).
\end{align}
\label{PTlm}%
\end{subequations}
Then the two sets of coefficients $(q_{lm},s_{lm},t_{lm})$ and $(\widetilde\Pol_{lm},\widetilde\Tor_{lm})$ are related by:
\begin{subequations}
\begin{align}
q_{lm} &= \frac{l(l+1)}{r}\, \widetilde\Pol_{lm}, \label{qlmPlm} \\
s_{lm} &= \sqrt{l(l+1)}\, \frac{1}{r}\dr{}\left(r\,\widetilde\Pol_{lm}\right), \label{slmPlm}\\
t_{lm} &= -\sqrt{l(l+1)}\, \widetilde\Tor_{lm}. \label{tlmTlm}
\end{align}
\label{qstPlm}%
\end{subequations}
It is clear from \Eq{qstPlm} that $q_{lm}$ and $s_{lm}$ are not independent but are related by
\begin{equation}
\frac{\sqrt{l(l+1)}}r s_{lm} = \dr{q_{lm}} + \frac2r q_{lm}.
\label{qlmslm}
\end{equation}
\Eq{qlmslm} can also be derived from $\nabla \cdot \V B=0$ by taking the divergence of \Eq{Bvsh}.

\section{Spectra of the scalar fields $\Pol$ and $\Tor$}
\label{FP_FT}

In the main text, various spectra for different vector fields are defined. In a very similar fashion, we can also consider the spectra for a scalar field. We define the spectrum $F_\Pol(l,r,t)$ for the poloidal potential $\Pol$ in \Eq{BPT2} by,
\begin{equation}
\sum_{l=1}^\infty F_\Pol(l,r,t) = \frac1{4\pi} \oint |\Pol|^2 \dOmega.
\end{equation}
Then using the expansion in \Eq{PTlm1}, we get
\begin{equation}
F_\Pol = \frac1{2l+1} \sum_{m=0}^l \big|\widetilde\Pol_{lm} \big|^2\, (4-3\delta_{m,0}).
\end{equation}
Similarly, we have for the spectrum of $\dot\Pol=\partial\Pol/\partial t$,
\begin{equation}
F_{\dot\Pol}(l,r,t) = \frac1{2l+1} \sum_{m=0}^l \big| \dot{\widetilde\Pol}_{lm}(r,t) \big|^2 (4-3\delta_{m,0}).
\label{svFPdef}
\end{equation}
Finally, the time-scale spectrum for $\Pol$ is defined as
\begin{equation}
\tau_\Pol(l,r) \equiv \avg{\! \sqrt{\frac{F_\Pol(l,r,t)}{F_{\dot\Pol}(l,r,t)}} \,}_{\!t}
= \avg{\! \sqrt{\frac{ \sum_m |\widetilde\Pol_{lm}|^2 }{ \sum_m |\dot{\widetilde\Pol}_{lm}|^2}} \,}_{\!t}.
\end{equation}
However because of \Eq{qlmPlm}, we see that $\tau_\Pol$ is identical to $\tau_r$ defined in \Eq{taurdef}. Following the same procedure, the time-scale spectrum for the toroidal potential $\Tor$ is given by:
\begin{equation}
\tau_\Tor(l,r) = \avg{\! \sqrt{\frac{ \sum_m |\widetilde\Tor_{lm}|^2 }{ \sum_m |\dot{\widetilde\Tor}_{lm}|^2}} \,}_{\!t}.
\end{equation}
And because of \Eq{tlmTlm}, $\tau_\Tor$ is identical to $\tauTor$ given in \Eq{tautordef}.

\label{lastpage}

\end{document}